\documentclass[11pt,preprint]{aastex}

\newcommand{\kB}{k_\mathrm{B}}	
\newcommand{\EF}{E_\mathrm{F}}	

\newcommand{\myunit}[1]{\mbox{$\,\mathrm{#1}$}}

\newcommand{\gram}{\myunit{g}}	
\newcommand{\cm}{\myunit{cm}}	
\newcommand{\second}{\myunit{s}} 
\newcommand{\K}{\myunit{K}}	
\newcommand{\dyne}{\myunit{dyne}} 
\newcommand{\erg}{\myunit{erg}} 

\newcommand{\fermi}{\myunit{fm}} 
\newcommand{\MeV}{\myunit{MeV}} 

\newcommand{\Msun}{\mbox{$M_\odot$}}

\newcommand{\yr}{\myunit{yr}}	
\newcommand{\km}{\myunit{km}}	

\newcommand{\ee}[1]{\times 10^{#1}}

\renewcommand{\EF}{E_{\rm F,e}}	

\newcommand{\eN}{\epsilon_N}	
\newcommand{\enu}{\epsilon_\nu}	
\newcommand{\ePhi}{e^{\Phi/c^2}}	
\newcommand{\ePhisq}{e^{2\Phi/c^2}} 
\newcommand{\eLambda}{(1-2G\mg/rc^2)^{-1/2}} 
\newcommand{\LAo}{L_A^\circ}	
\newcommand{\Tcore}{T_{\rm core}} 
\newcommand{\GTcore}{T_{\rm core,9}} 
\newcommand{\avmod}{AV18+$\delta$v+UIX*}
\newcommand{\avold}{AV14+UVII}
\newcommand{\mg}{m}
\newcommand{\Mg}{\mathcal{M}}
\newcommand{\modelone}{M1.4-14}
\newcommand{\modeltwo}{M1.8-14}
\newcommand{\modelthree}{M1.4-18}
\newcommand{\modelfour}{M1.8-18}

\begin{document}

\slugcomment{To be published in \emph{The Astrophysical Journal}}

%
%

\title{
   Nuclear heating and melted layers in the inner crust of an accreting
   neutron star
}

\author{
   Edward F. Brown\altaffilmark{1}
}

\affil{
   Department of Physics and Department of Astronomy
}
\affil{
   601 Campbell Hall \# 3411, University of California, Berkeley, CA
   94720--3411
}

\altaffiltext{1}{
   Current Address: University of Chicago, Astronomy and
   Astrophysics, 5640 S Ellis Ave, Chicago, IL 60637;
   brown@oddjob.uchicago.edu
}

\email{
   brown@oddjob.uchicago.edu
}

\shortauthors{Brown}
\shorttitle{Deep heating in accreted neutron star crusts}
\begin{abstract}

A neutron star in a long-lived, low-mass binary can easily accrete
enough matter to replace its entire crust.  Previous authors noted that
an accreted crust, being formed from the burning of accreted hydrogen
and helium, allows a series of non-equilibrium reactions, at densities
$\gtrsim 6\ee{11}\gram\cm^{-3}$, which release a substantial amount of
heat ($\sim 1\MeV$ per accreted nucleon).  Recent calculations by Schatz
et al.\ showed that the crystalline lattice of an accreted crust is also
likely to be quite impure.  This paper discusses the thermal structure
of such a neutron star and surveys how the crust reactions and
impurities affect the crust temperature.  During accretion rapid enough
to make the accreted hydrogen and helium burn stably ($\dot{M}\sim
10^{-8}\Msun\yr^{-1}$; typical of the brightest low-mass neutron star
binaries), most of the heat released in the crust is conducted into the
core, where neutrino emission regulates the temperature.  As a result
there is an inversion of the thermal gradient: the temperature decreases
with depth in the inner crust.  The thermal structure in the crust at
these high accretion rates is insensitive to the temperature in the
hydrogen/helium burning shell.  When the crust is very impure, the
temperature can reach $\approx 8\ee{8}\K$ at densities $\gtrsim
6\ee{11}\gram\cm^{-3}$.  This peak temperature depends mostly on the
amount of heat released and the thermal conductivity and in particular
is roughly independent of the core temperature.  The high crust
temperatures are sufficient to melt the crystalline lattice in thin
layers where electron captures have substantially reduced the nuclear
charge.

\end{abstract}


\section{Introduction}
\label{s:introduction}

Many of the known neutron stars reside in low-mass x-ray binaries.
These sources typically accrete at rates of $10^{-11}\Msun\yr^{-1}$ to
$10^{-8}\Msun\yr^{-1}$ and show no conclusive evidence, such as
cyclotron lines or coherent pulsations (in the persistent emission), of
a magnetic field.  In contrast to studies of isolated neutron star
cooling \citep[for a review, see][]{tsuruta98}, there has been much less
interest in the interior thermal state of an accreting neutron star.
Originally, studies of how accretion affected a neutron star's thermal
structures were motivated by the challenge of explaining the type I
x-ray bursts of some of these sources.  Both \citet{lamb78:_nuclear_x}
and \citet{ayasli82} estimated the steady-state core temperature by
balancing the heating from hydrogen and helium shell burning with
neutrino and radiative losses.  Later, \citet{fujimoto84} and
\citet{hanawa84} calculated the thermal evolution of the entire neutron
star, both for steady hydrogen and helium burning \citep{fujimoto84} and
for repeated shell flashes \citep{hanawa84}.  In all of these works, the
only heat sources considered were the hydrogen/helium burning and the
influx of entropy by the accreted matter.  Both \citet{fujimoto84} and
\citet{hanawa84}, who considered accretion rates $\lesssim
5\ee{-9}\Msun\yr^{-1}$, found that the deep crust and core would
gradually become isothermal, at a temperature $\gtrsim 10^8\K$, if there
were no enhanced neutrino cooling; otherwise, the core would remain
chilled at temperatures $\sim 10^7\K$.  Without heat sources in the
crust, the temperature of the deep crust tracks that of the core and is
therefore sensitive to the cooling from neutrino processes active in the
core.

Unlike an isolated neutron star, the crust of an accreting neutron star
is not in statistical nuclear equilibrium, but rather has a composition
set by the nuclear history of the accreted material
\citep{sato79,blaes90:_slowl,blaes92,haensel90b,haensel90a}.  The
atmosphere is composed of the accreted helium and hydrogen and any
metals present (see \citealt*{bildsten92} for a discussion).  This
accumulated fuel eventually burns to heavier elements.  The accretion of
fresh fuel shoves the original crust deeper and, if continued over a
long enough interval, will eventually replace the original crust with
one formed by the ashes of hydrogen/helium burning.  Compression of the
crust by the weight of continually accreting material induces
non-equilibrium reactions that release heat.

The composition of the replaced crust is uncertain.  Improved treatments
of the physics of hydrogen and helium nuclear burning revealed that the
ashes of this burning are unlikely to be a pure species, e.g., iron.
The mixture is formed by the rp-process
\citep{wallace81:_explos,champagne92:_explos,van94:_react,schatz98}, a
sequence of rapid proton captures onto seed nuclei provided by helium
and CNO burning.  Calculations of the nucleosynthetic yield from the
rp-process have been done, both for unstable burning during an x-ray
burst \citep{koike99} and for steady-state hydrogen and helium burning
\citep{schatz99}.  The ashes of the stable burning are a motley mix of
iron-peak elements, so that the crust formed from these ashes will
likely be very impure \citep{schatz99}.

This paper studies the crustal temperatures of steadily accreting
neutron stars with low magnetic fields.  There are two differences with
\citet*{miralda-escude90} and \citet{zdunik92}, both of which included
crust reactions in the neutron star's thermal balance.  First, this work
considers the stable regime of hydrogen/helium burning, which requires
rapid accretion (near the Eddington limit, $\sim 10^{-8}\Msun\yr^{-1}$).
At the low accretion rates ($\dot{M}\lesssim 10^{-10}\Msun\yr^{-1}$,
roughly two orders of magnitude less than the Eddington limit)
considered by \citet{miralda-escude90} and \citet{zdunik92}, the crust
is basically isothermal, with a temperature locked to that of the core.
Second, this work allows for an impure crust by surveying both high- and
low-conductivity cases.  Previous calculations have assumed the impurity
concentration to be much less than unity.

At the rapid accretion rates considered here, the neutrino luminosity
from modified Urca processes and crust bremsstrahlung is significant and
causes the temperature to decrease with depth in the inner crust
\citep{brown98a}.  Almost all of the heat produced in the crust flows
inward.  Moreover, the reduced conductivity of the impure crust produces
a peaked thermal profile with a maximum temperature where the nuclear
reactions heat the crust, at densities $\gtrsim 6\ee{11}\gram\cm^{-3}$.
The thermal profile in the crust is primarily determined by the ability
of the inner crust to conduct a flux of $\sim 1\MeV$ per accreted baryon
into the core, and is insensitive to the temperature of the
hydrogen/helium burning shells.  If the crust is very impure, the crust
reaches temperatures $\approx 8\ee{8}\K$; the value of this temperature
only weakly depends on the core temperature.

Electron captures in the crust reduce the charge of the nuclei ($Z$) and
hence the electrostatic binding of the lattice.  For the hottest
temperatures in the crust, this low-$Z$ lattice melts where the charge
is lowest ($Z\lesssim 15$ for the composition of \citealt{haensel90b}).
As a result, the inner crust of the neutron star comes to resemble a
``layer cake,'' with alternating layers of lattice and liquid.

This paper is relevant for the brightest low-mass x-ray binaries.  These
weakly magnetized neutron stars are considered possible progenitors of
millisecond pulsars \citep[for a review, see][]{bhattacharya95}, and
there has been much theoretical interest in the evolution of the crust
magnetic field (\citealt{romani90}; \citealt{geppert94};
\citealt{urpin95}; \citealt{konar97}; \citealt{brown98a};
\citealt*{urpin98a}).  Many of these neutron stars rotate within an
apparently narrow range of spin frequencies $\sim 300\rm\,Hz$
\citep[e.g.]{vdKlis98}.  One possibility for this convergence of spin
frequencies is that gravitational radiation from the neutron star
balances the accretion torque
\citep{bildsten98:gravity-wave,andersson99}.  The source for the
gravitational radiation could be a mass quadrupole formed by misaligned
electron capture layers in the crust \citep{bildsten98:gravity-wave} or
a current quadrupole from an r-mode instability in the core
(\citealt{andersson98:_new_class}; \citealt{friedman98:_axial_instab};
\citealt*{lindblom98:_gravit_radiat_instab}; \citealt{owen98:_gravit};
\citealt*{andersson99:_gravit_radiat}; \citealt*{andersson99}).  All of
these problems depend on the thermodynamics of the neutron star's crust
and core and motivate this paper.

\subsection{An overview of the problem}
\label{sec:An-overview-problem}

A neutron star has several distinctive regions.  The \emph{core}
consists of uniform $npe^-$ (in the least dense parts).  At a baryon
density less than $n \approx 0.6 n_s$, where $n_s=0.16\fermi^{-3}$ is
the saturation density\footnote{%
   Density is measured in units of $\fermi^{-3}$ and pressure in units
   of $\MeV\fermi^{-3}$.  Note that throughout the crust, $\rho\approx
   n\, m_u=1.66\ee{14}(n/0.1\fermi^{-3})\gram\cm^{-3}$, where
   $m_u=1.66\ee{-24}\gram$ is the atomic mass unit, and
   $1\MeV\fermi^{-3}=1.6\ee{33}\dyne\cm^{-2}$.}
of nuclear matter, individual nuclei appear \citep*{pethick95}.  The
portion of the neutron star exterior to this point, the \emph{inner
crust}, is composed of nuclei, degenerate neutrons, and relativistic
degenerate electrons.  Where the electron Fermi energy is less than
about twice the nuclear bulk energy ($\approx 30\MeV$; see
\citealt{pethick98}), free neutrons can no longer exist in
$\beta$-equilibrium.  This point, \emph{neutron drip}, has a density
$\approx 0.0023 n_s$ ($4\ee{-4}\fermi^{-3}$) and marks the boundary
between the inner and the \emph{outer crust}.  At lesser densities, the
crust is made of nuclei and electrons, with the degenerate electrons
supplying the pressure.  The outer and inner crust collectively occupy
the outermost kilometer or so of the neutron star and contain a total
mass of order $0.01 \Msun$.

\begin{figure}[tp]
\centering{\includegraphics[width=88mm]{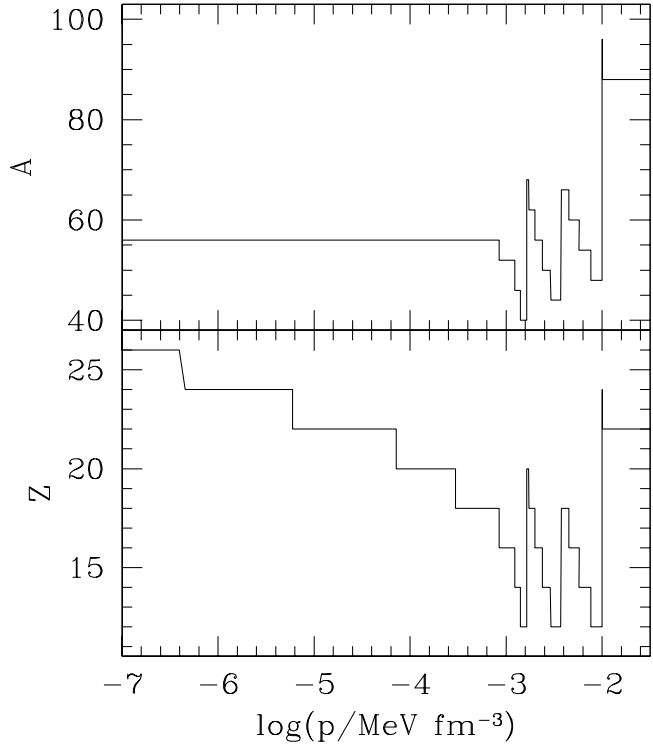}}
\caption{\label{fig:crust-schematic}%
   Composition of an accreted crust, as computed by
   \protect\citet{haensel90a}.  Shown are the mass number (\emph{top
   panel}) and charge number (\emph{bottom panel}) as a function of
   pressure.  The mass resting above a given isobar is given by equation
   (\protect\ref{eq:M-P}).  Electron captures produce a downward jump in
   $Z$; neutron emissions produce a downward jump in $A$.  The upward
   jumps in both $Z$ and $A$ indicate a pycnonuclear reaction.
}
\end{figure}

The boundaries of the crust are demarcated by surfaces of constant
pressure.  For a thin crust, the mass above a given isobar is fixed by
the surface gravity and area.  As a consequence, accretion during a time
brief compared with $M/\dot{M}$ (so that the overall structure of the
neutron star remains roughly constant) pushes the underlying crust
through these compositional boundaries.  Low-mass x-ray binaries live
for more than $10^8\yr$ \citep*{webbink83}; even at
$\dot{M}=10^{-10}\Msun\yr^{-1}$, the neutron star can easily accrete
enough material from the secondary to replace its entire crust.  This
replaced crust is composed of the ashes of hydrogen and helium burning
and is quite different in composition from the original.  Unlike during
the neutron star's hot birth, the crust does not burn to nuclear
statistical equilibrium.  As the ashes of the hydrogen and helium
burning are pushed deeper into the crust, the rising electron Fermi
energy induces a series of electron captures
\citep{haensel90a,sato79,blaes90:_slowl}.  Further compression of this
low-$Z$ material causes neutron emissions and pycnonuclear reactions
\citep{haensel90a,sato79}.  A schematic of the composition, from the
calculation of \citet{haensel90a}, is shown in
Figure~\ref{fig:crust-schematic}.  Each decrease in the nuclear charge
$Z$ (\emph{bottom panel}) is from an electron capture, and each decrease
in the nuclear mass number $A$ (\emph{top panel}) is from a neutron
emission.  Where a pycnonuclear reaction occurs, both $Z$ and $A$
double.

The pycnonuclear reactions and neutron emissions liberate $E_N\approx
1\MeV$ for each baryon accreted and heat the crust at a rate $L_N
\approx 10^{36} (\dot{M}/10^{-8}\Msun\yr^{-1})\erg\second^{-1}$.  To set
the scale for the core temperature, note that the core must be at a
temperature $\approx 4\ee{8} (L_N/10^{36}\erg\second^{-1})^{1/8}\K$ for
modified Urca processes \citep{friman79,yakovlev95} to radiate a
neutrino luminosity equal to $L_N$.  If the nucleons in the core were
superfluid and the modified Urca processes correspondingly suppressed,
crust neutrino bremsstrahlung (\citealt{maxwell79:_neutr};
\citealt{yakovlev96}; \citealt*{haensel96b}; \citealt{itoh96:_neutr})
can also balance $L_N$ for an inner crust temperature $\approx
6\ee{8}(L_N/10^{36}\erg\second^{-1})^{1/6}\K$.

The luminosity from the steady-state hydrogen/helium burning ($\approx
5\MeV$ per accreted baryon; \citealt{schatz99}) is much larger than that
flowing out from deeper in the crust.  As a result, the temperature at
the base of the hydrogen/helium burning shell is determined by the
luminosity there \citep[for a review, see][]{bildsten98:_nuclear} and is
$\approx 5\ee{8}\K$ for $\dot{M}\approx 10^{-8}\Msun\yr^{-1}$
\citep{brown98a,schatz99}.  The conductivity is much greater in the
inner crust (the conductivity in the crust increases with density),
while the thickness of the inner crust is only a factor of two greater
than the thickness of the outer crust.  For a similar thermal gradient,
the inner crust can carry a much larger flux. If the change in
temperature between the hydrogen/helium burning shell and neutron drip
is of the same order as the change in the inner crust, than most of the
nuclear luminosity generated in the crust will flow into the core
\citep{brown98a}.

\subsection{Outline of this paper}
\label{sec:outline-this-paper}

The remainder of this paper paints in the crude picture just sketched.
Sections~\ref{sec:hydro_structure} and~\ref{sec:Thermal-Structure}
develop the details of the calculation, which proceeds in two steps.
First, the hydrostatic structure of the neutron star is computed for
different equations of state (\S~\ref{sec:equation-state}) and neutron
star masses.  These hydrostatic models serve as background for the
thermal computations, which are discussed in
\S~\ref{sec:Thermal-Structure}.  Section~\ref{sec:Thermal-Structure}
also describes the relevant microphysics: the heating by crust reactions
and the cooling by the crust and core neutrino emissivity
(\S~\ref{sec:nucl-heat-neutr}); the reduction of the core neutrino
emissivity by neutron and proton superfluidity
(\S~\ref{sec:Superfluidity}); and the conductivity of the crust and core
(\S~\ref{sec:heat-transport}), including the effect of impurities.

Section~\ref{s:results} contains the results of these calculations,
which are split into three parts.  First, there is discussion on the
nature of the thermal profile and its dependence on the crust
composition and the core neutrino emissivity
(\S~\ref{sec:infl-micr-therm}).  Section~\ref{sec:Simple-expr-crust}
presents some analytical formulae for the crust temperature; these
formula are used (\S~\ref{sec:approach-to-lower}) to show how the
high-accretion rate solutions discussed here connect with those at lower
accretion rates \citep[e.g.,][]{miralda-escude90,zdunik92}.  The melting
and refreezing of the crystalline lattice in the inner crust are
discussed in \S~\ref{sec:Crust-melting}.

\section{Hydrostatic structure}\label{sec:hydro_structure}

Throughout the crust and core, the pressure is supplied by degenerate
particles with Fermi energies $\gg \kB T$, and so the equation of state
(EOS) scarcely depends on temperature.  The crust reactions heat the
core on a timescale
\begin{equation}
\label{eq:heating-timescale}
   \tau_H \sim \left(\frac{M}{m_u}\right) \frac{CT}{L_N} \approx
   \frac{M}{\dot{M}}\frac{CT}{E_N}\ll\frac{M}{\dot{M}}.
\end{equation}
In this equation $C$ is the specific heat per baryon, $M/m_n$ is
approximately the number of baryons in the star, and $M/\dot{M}$ is the
timescale for the mass to increase from accretion.  If the heat is
stored in the electrons (as would be the case if the neutrons were
superfluid), the heat content per baryon
\citep[e.g.,][]{landau80:_statis_physic} is $CT \approx \pi^2 \kB T (\kB
T/\EF)\ll E_N$, where $\EF$ is the electron Fermi energy.  The heat
content per baryon is similar if the neutrons are normal
\citep{lamb78:_nuclear_x}.  Equation~(\ref{eq:heating-timescale}) shows
that over timescales necessary to establish a thermal steady-state, the
mass of the star changes only slightly, and so the hydrostatic equations
need not be solved simultaneously with the thermal equations.  Using a
fixed hydrostatic structure simplifies the thermal calculation.

To calculate the temperature as a function of radius requires
integrating the heat transport equations over the star.  The strong
gravitational field modifies the heat flow.  In an isothermal star,
where there is no heat flow, the \emph{redshifted} temperature is
constant, while the proper (as measured by a local thermometer)
temperature increases as one moves toward the stellar center.  Because
the neutrino emissivity is a strong function of temperature, the thermal
transport equations must account for gravitational effects.  The
appropriate equations, solved for stellar mass and EOS, are the
post-Newtonian stellar structure equations \citep{thorne77} for the
radius, gravitational mass, potential, and pressure:
\begin{eqnarray}
\label{e:radius}
   \frac{\partial r}{\partial a} &=& \left(4\pi r^2 n\right)^{-1}
   \left(1-\frac{2G\mg}{rc^2}\right)^{1/2} \\
\label{e:mass}
   \frac{\partial \mg}{\partial a} &=& \frac{\rho}{n}
   \left(1-\frac{2G\mg}{rc^2}\right)^{1/2} \\
\label{e:potential}
   \frac{\partial \Phi}{\partial a} &=& 
   \frac{G\mg }{4\pi r^4 n} \left(1+\frac{4\pi r^3 p}{\mg
   c^2}\right) \left(1-\frac{2G\mg}{rc^2}\right)^{-1/2}\\ 
\label{e:pressure}
   \frac{\partial p}{\partial a} &=& -\frac{G\mg}{4\pi r^4}
   \frac{\rho}{n}
   \left(1+\frac{p}{\rho c^2}\right) 
   \left(1+\frac{4\pi r^3 p}{\mg c^2}\right)
   \left(1-\frac{2G\mg}{rc^2}\right)^{-1/2} .
\end{eqnarray}
In these equations the Lagrangian variable $a$ is the total number of
baryons inside a sphere of area $4\pi r^2$, and $\rho$ is the mass
density.  The potential $\Phi$ appears in the time-time component of the
metric as $e^{\Phi/c^2}$ (it governs the redshift of photons and
neutrinos; \citealt*{misner73:_gravit}) and satisfies the boundary
condition that at the stellar surface $e^{2\Phi/c^2}|_{r=R} =
1-2G\Mg/Rc^2$, where $\Mg$ and $4\pi R^2$ are the total gravitational
mass and surface area of the neutron star.

\subsection{Equation of state}\label{sec:equation-state}

For purposes of calculating the crust EOS, the ashes of hydrogen and
helium burning are presumed to be pure iron (but see the discussion in
\S~\ref{sec:heat-transport} on how the composition affects the energy
release and heat transport).  As a mass element is compressed to greater
densities and pressures, the rising electron Fermi energy triggers a
series of electron captures, neutron emissions, and pycnonuclear
reactions \citep{sato79,blaes90:_slowl,haensel90a}.  At any given
density, only one species is assumed present (see
Figure~\ref{fig:crust-schematic}) according to the composition
calculated by \citet{haensel90b,haensel90a}.

In the outer crust, relativistic degenerate electrons of density
$n_e=Y_e n$ supply the pressure.  The electron chemical potential is
basically the Fermi energy $\EF=m_e c^2[1+(3\pi^2
n_e)^{2/3}\lambda_e^2]^{1/2}$, where $\lambda_e=386.2\fermi$ is the
electron Compton wavelength.  I calculate the electron pressure, which
is approximately $n_e\EF/4$, from the interpolation formula of
\citet{paczynski83}.  The lattice pressure is calculated from the ionic
free energy, which is a function of
\begin{equation}
\label{eq:Gamma}
   \Gamma = \frac{Z^2 e^2}{\kB T}\left(\frac{4\pi}{3}n_N\right)^{1/3},
\end{equation}
where $n_N$ is the density of nuclei.  I use the fits of
\citet{farouki93} to Monte-Carlo simulations of the free energy.  (In
the crust, the free energy per nucleus is to lowest order just the
Madelung energy, $\approx -0.9 \Gamma \kB T$.)  These fits are valid for
$\Gamma > 1$, which is always the case for the density-temperature
regime of interest.  Following \citet{farouki93}, I presume the nuclei
are crystalline for $\Gamma\ge 173$.  The binding energy of the nuclei
are computed from a compressible liquid-drop model \citep{mackie77}.
This formula accounts for an external neutron gas and is therefore
applicable at densities greater than neutron drip.  The energy density
and pressure of the neutron gas (which differ from that of an ideal
degenerate gas) are also computed with this model in the limit of a
vanishing proton fraction.

\begin{figure}[tp]
\centering{\includegraphics[width=88mm]{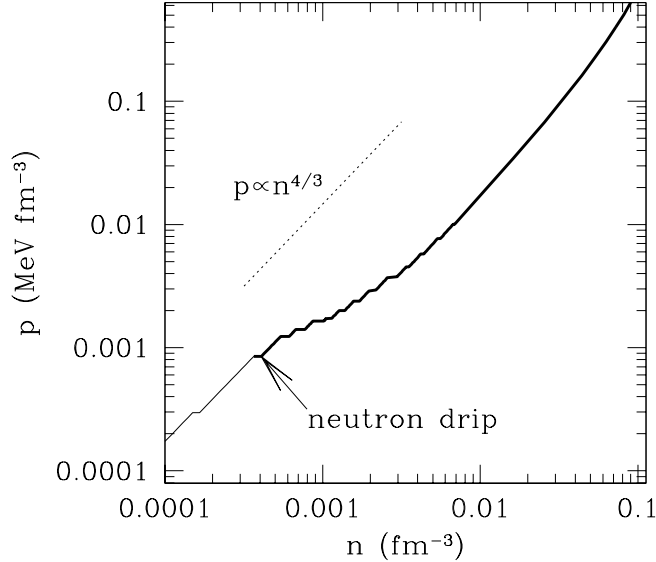}}
\caption{\label{fig:crustEOS}%
   Pressure as a function of baryon density $n$ in the crust.  Each
   electron capture is visible as a discontinuity in $n$.  Where free
   neutrons are present (\emph{heavy solid line}) the pressure deviates
   from the piecewise scaling for degenerate, relativistic electrons
   (\emph{dotted line}).
}
\end{figure}

Summing the pressure contributions from electrons, ions, and neutrons
gives the crust EOS, which agrees with that of \citet{haensel90b} to the
accuracy of their table.  Figure~\ref{fig:crustEOS} displays this
relation, $p(n)$, throughout the inner crust.  For reference, the
$p\propto n^{4/3}$ relation appropriate for an EOS dominated by
relativistic, degenerate electrons is also shown (\emph{dotted line}).
Free neutrons are present (\emph{heavy solid line}) for
$n>3.6\ee{-4}\fermi^{-3}$.  As noted by \citet{haensel90b}, for
$n>0.04\fermi^{-3}$, the free neutrons provide most of the pressure, and
the ionic composition becomes less and less important to the EOS.  In
this regime I use the $p(n)$ fit of \citet{negele73}.

At $n\gtrsim 0.1\fermi^{-3}$, the nuclei dissolve into uniform nuclear
matter \citep{pethick95}.  I select two sample core equations of state
for comparison.  The first EOS is a fit \citep{lai94} to the \avold\
interaction, which is the Argonne V14 potential, with a three-nucleon
interaction prescribed by the Urbana VII potential \citep*{wiringa88}.
The second EOS, called \avmod, is a Skyrme-type Hamiltonian fit
\citep*[appendix A]{akmal98} to the Argonne V18 potential, with
relativistic boost corrections and the three-nucleon interaction UIX*
\citep{akmal98}.  The components of both interactions are neutrons,
protons, electrons, and, where $\EF>m_\mu c^2=105.66\MeV$, muons.  I do
not consider, for simplicity, other possible components (e.g., hyperons
or quark matter) in the EOS.  To construct a table suitable for
interpolating $n(p)$, I calculate for each $n$ the proton fraction
$Y_p=n_p/n$ and electron fraction $Y_e=n_e/n$ from the equations for
$\beta$-equilibrium, $\mu_n-\mu_p = \mu_e = \mu_\mu$, and charge
neutrality, $n_p=n_e+n_\mu$.  Given $(Y_p,Y_e)$, I then compute the mass
density $\rho$ and pressure $p=c^2(-\rho+n\partial\rho/\partial n)$.

There have been many attempts to calculate the density of the phase
transition from the inner crust to the core \citep[see][and references
therein]{pethick95:_matter}.  I adopt the following approach.  The
density and pressure of the \avmod\ EOS equal those of \citet{negele73}
at $n = 0.078\fermi^{-3}$, $p=0.39\MeV\fermi^{-3}$.  I therefore take
this density as the transition from crust to core; there is no density
discontinuity in this case.  For the \avold\ EOS, the energy density is
always greater than that of \citet{negele73}, and so I choose the
maximum crust density to be $0.1\fermi^{-3}$ ($p=0.60\MeV\fermi^{-3}$).
In this case, there is a substantial density jump (from
$n=0.1\fermi^{-3}$ to $n=0.13\fermi^{-3}$) between crust and core.  The
choice $n=0.1\fermi^{-3}$ as the upper limit for the crust density
reflects recent detailed calculations \citep{pethick95} of the phase
transition.

For equilibrium crust compositions, it becomes energetically favorable
for nuclei to turn inside-out in the inner crust and form a phase with
bubbles of neutron gas encased in bound nuclear matter
(\citealt*{lorenz93:_neutr}; \citealt{oyamatsu93:_nuclear}).  Because
the charge of nuclei in an accreted crust is less than that of the
equilibrium composition, it is possible that the nuclei do not turn
inside-out.  The condition for this inversion \citep[see,
e.g.,][]{pethick95:_matter} is that the nuclear radius be more than
one-half the Wigner-Seitz radius, $(4\pi n_N/3)^{-1/3}$.  For the core
EOS \avmod, this ratio at the bottom of the crust is (for $Z=20$,
$A=100$, $Y_n=0.8$, where $Y_n=n_n/n$ is the neutron fraction) 0.50.
For the core EOS \avold, the ratio is 0.54.  In the absence of a more
detailed calculation of the composition, I am unable to say if an
accreted crust contains non-spherical nuclei and have not explored this
possibility.

\subsection{Construction of models}

With $n(p)$ specified by interpolation from a table, I integrate
equations (\ref{e:radius})--(\ref{e:pressure}) with a fourth-order
Runge-Kutta integration algorithm \citep{pre92}.  Borrowing a technique
used by \citet{vanRiper91}, the code restricts the stepsize $\Delta a$
to be always less than some fraction $f$ of the reciprocal sum of the
radial and baric scale heights,
\begin{equation}
\label{eq:stepsize}
   \Delta a \le f\left(\frac{\partial\ln r}{\partial a} -
    \frac{\partial\ln P}{\partial a}\right)^{-1}.
\end{equation}
Starting from a fixed central pressure, I expand the hydrostatic
equations about the center $a=\mg=r=0$ and integrate outwards until the
pressure is less than $3.4\ee{-8}\MeV\fermi^{-3}$ (corresponding to a
density $n\approx 2.4\ee{-8}\fermi^{-3}$, about a factor of 10 greater
than where the helium burning ends; \citealt{schatz99}).  At this
pressure, the radius and mass are constant to within $10^{-5}R$ and
$10^{-9}M$, respectively.  The algorithm iteratively adjusts the central
pressure until a target gravitational mass $\Mg$ is reached.  The
integration steps are then stored for later use in solving the thermal
equations.  For each of the two equations of state, \avold\ and \avmod,
I compute two masses, $\Mg=1.4\Msun$ and $\Mg=1.8\Msun$; a summary of
these four structures is provided in Table \ref{t:models}.

\begin{deluxetable}{clllllll}
\tablewidth{0pt}
\tablecaption{Hydrostatic structures\label{t:models}}
\tablehead{%
   \colhead{model} &
   \colhead{$M$} &
   \colhead{$\Mg$} & 
   \colhead{$R$} &
   \colhead{$1+z$} &
   \colhead{$g$} &
   \colhead{$p_{\rm crust}$} &
   \colhead{$\Delta M$} \\
   & 
   \colhead{($\Msun$)} & 
   \colhead{($\Msun$)} &
   \colhead{(km)} &
   &
   \colhead{($10^{14}\cm\second^{-2}$)} &
   \colhead{($\MeV\fermi^{-3}$)} &
   \colhead{($\Msun$)}
}
\startdata
\modelone\tablenotemark{a}  & 1.58 & 1.4 & 10.7 & 1.28 & 2.08 & 0.60 & 0.025\\
\modeltwo\tablenotemark{a}  & 2.12 & 1.8 & 10.4 & 1.43 & 3.13 &      & 0.017\\
\modelthree\tablenotemark{b}& 1.56 & 1.4 & 11.6 & 1.25 & 1.74 & 0.39 & 0.021\\
\modelfour\tablenotemark{b} & 2.09 & 1.8 & 11.3 & 1.38 & 2.61 &      & 0.015\\
\enddata
\tablenotetext{a}{%
   EOS is a fit \protect\citep{lai94} to \avold\
   \protect\citep{wiringa88}}
\tablenotetext{b}{%
   EOS is \avmod\ \protect\citep[appendix A]{akmal98}}
\tablecomments{%
   $M$ is the number of baryons in the star, multiplied by
   $m_n=1.67\ee{-24}\gram$; $\Mg$ is the gravitational mass of the star;
   $p_{\rm crust}$ is the pressure at the base of the crust; and $\Delta
   M$ is the number of baryons in the crust multiplied by $m_n$.
}
\end{deluxetable}

Despite continuous accretion, the pressure is a good Eulerian coordinate
throughout the crust \citep{bildsten92,brown98a}, and I shall plot the
temperature and luminosity against it.  The mass contained in the crust
above a given isobar is to lowest order (from expanding
eq.~[\ref{e:pressure}]; \citealt{lorenz93:_neutr})
\begin{eqnarray}
   \Delta M &=& m_n \Delta a \approx \frac{p}{g}4\pi R^2 \nonumber\\
      &=& 0.05\Msun \left(\frac{p}{\MeV\fermi^{-3}}\right)
      \left(\frac{R}{10\km}\right)^2
      \left(\frac{2\ee{14}\cm\second^{-2}}{g}\right),
\label{eq:M-P}
\end{eqnarray}
where $g=G\Mg (1+z)/R^2$ is the gravitational acceleration and
$(1+z)=(1-2G\Mg/Rc^2)^{-1/2}$ is the surface redshift.
Table~\ref{t:models} lists $\Delta M$, $g$, and $1+z$ for the two masses
and two EOSs considered in this paper.

\section{Thermal Structure}
\label{sec:Thermal-Structure}

With a hydrostatic structure specified, the luminosity $L$ and
temperature $T$ are found by solving the entropy and flux equations
\citep{thorne77},
\begin{eqnarray}
\label{eq:entropy}
   e^{-2\Phi/c^2} \frac{\partial}{\partial r}\left(L \ePhisq\right)
      - 4\pi r^2 n \left(\eN - \enu\right)\eLambda &=& 0\\
\label{eq:flux}
   e^{-\Phi/c^2} K \frac{\partial}{\partial r}\left(T \ePhi\right) 
      + \frac{L}{4\pi r^2}\eLambda &=& 0.
\end{eqnarray}
Here $\eN$ and $\enu$ are the nuclear heating and neutrino emissivity
per baryon, and $K$ is the thermal conductivity.  I neglect in
equation~(\ref{eq:entropy}) terms arising from compressional heating, as
they are of order $T\Delta s (\dot{M}/M)$ \citep{fujimoto82:_helium},
$s$ being the specific entropy, and are negligible throughout the
degenerate crust and core \citep{brown98a}.  The physics of the problem
is contained in the heating $\eN$, neutrino cooling $\enu$, and the
conductivity $K$; the following sections discuss each in turn.

\subsection{Nuclear heating and neutrino cooling}
\label{sec:nucl-heat-neutr}

As mentioned in \S~\ref{sec:equation-state}, the crust electron captures
reduce the charge of the nuclei enough to trigger pycnonuclear
reactions.  The rate of pycnonuclear reactions is governed by the supply
of low-Z nuclei, which is in turn determined by the rate of the
preceding electron capture.  As a result, even though the pycnonuclear
reactions are better described at typical crust temperatures as
strongly screened fusion reactions \citep{salpeter69:_nuclear}, they are
insensitive to temperature and hence not susceptible to a thermal
instability.

To find the overall thermal profile, I do not need to resolve the
individual capture layers.  Rather, I distribute the reaction heat per
baryon, $E_N=1\MeV$, over a pressure interval $\Delta p$ between
$8.7\ee{-4}\MeV\fermi^{-3}$ and $3.4\ee{-2}\MeV\fermi^{-3}$ (this covers
the region where the pycnonuclear reactions occur).  The total nuclear
luminosity is $L_N=\dot{M}e^{-\Phi} E_N/m_n \approx L_A/200$, where
$L_A=\dot{M}c^2 z/(1+z) \approx GM\dot{M}/R$ is the accretion luminosity
(see \S~\ref{sec:Bound-cond-meth}) and $\dot{M}e^{-\Phi}$ is the
accretion rate as measured in the crust.  The heating term in
equation~(\ref{eq:entropy}) is therefore
\begin{equation}
\label{eq:nuclear-heating}
   4\pi r^2 n \eN = \left( \frac{E_N}{m_n} \dot{M} e^{-\Phi/c^2}
      \right) \left( \frac{1}{\Delta p} \frac{\partial p}{\partial r}
      \right),
\end{equation}
where $\partial p/\partial r$ is the Jacobean.

The cooling terms in equation~(\ref{eq:entropy}) are evaluated from
various fits to microscopic calculations.  Throughout much of the crust,
the dominant neutrino emissivity is neutrino pair bremsstrahlung,
$e^-+(A,Z)\to e^- + (A,Z)+\nu\bar{\nu}$
\citep{maxwell79:_neutr,haensel96b,itoh96:_neutr}.  Where the ions are
crystallized, the bremsstrahlung rates are exponentially suppressed
because the separation between electron energy bands is of order
$1\MeV\gg \kB T$ \citep{pethick94,pethick97}.  I use the fits of
\citet{haensel96b} for the emissivity where the ions are liquefied and
the fits of \citet{yakovlev96}, which include this suppression, where
the ions are crystallized.  If the crust is very impure (see the
discussion in \S~\ref{sec:heat-transport}), then the crust
bremsstrahlung will be dominated by electron-impurity scattering
\citep{pethick97}.  For $T\gtrsim 10^9\K$ and densities where
$\hbar\omega_{pe}/\kB T\gtrsim 1$, $\hbar\omega_{pe}\approx
0.056(\EF/1\MeV)\MeV$ being the electron plasma frequency, the plasma
neutrino process \citep{schinder87,itoh96:_neutr} becomes important.

This paper assumes a standard core neutrino emissivity, for which
modified Urca processes \citep{friman79,yakovlev95} dominate.  The phase
space available for scattering is strongly restricted if the nucleons
are superfluid and reduces the emissivity roughly as $\exp(-T/T_c)$
\citep{yakovlev95} where $T_c$ is the superfluid transition temperature.
In general, both the neutrons and protons must be superfluid to
substantially reduce the modified Urca neutrino luminosity.  The proton
modified Urca branch ($p + p \to p + n + e^+ + \nu_e$) is nearly as
efficient as the neutron branch \citep{yakovlev95}, and so a slight
increase in temperature is sufficient to compensate for the suppression
of just one of the modified Urca branches.

\subsection{The superfluid transition temperatures}
\label{sec:Superfluidity}

In the core, both the protons and neutrons are expected to be superfluid
over some range of densities (\citealt*{baym69:_super};
\citealt{hoffberg70}; \citealt{takatsuka93};
\citealt{amundsen85:triplet}; \citealt{amundsen85:singlet};
\citealt{elgaroey96}).  At lower densities, the neutrons pair in a
singlet ($^1S_0$) state, but at higher densities the repulsive core of
the interaction forces the neutrons to pair in a triplet ($^3P_2$)
state.  The protons in the core are expected to be in a $^1S_0$ state.
There is at present little agreement on the range of densities for which
the protons and neutrons are superfluid and on their transition
temperatures $T_c$ \citep[for a review, see][]{pethick95}.  The early
calculation of \citet{hoffberg70} for pure neutron matter found peak gap
energies of $1.6\MeV$ (singlet) and $\sim 5\MeV$ (triplet).  They also
found that the transition temperature remained high over a large range
of densities, implying that the entire core would be superfluid.  More
recent calculations of the neutron triplet pairing
\citep{takatsuka93,amundsen85:triplet} find a lower transition
temperature $\kB T_c^{\rm max}\lesssim 0.1\MeV$.  In addition, the range
of densities for which pairing occurs is restricted.
\citet{elgaroey96}, using a meson-exchange model for $\beta$-stable
matter, found that maximum gap energy was $\approx 0.018\MeV$ and that
the range of densities was restricted to $n \lesssim 0.13\fermi^{-3}$,
so that almost all of the core would be normal.  In all cases the proton
pairing gaps are somewhat larger, with $\kB T_c^{\rm max}\sim 1.0\MeV$,
and extend to densities several times the saturation density,
$n_s=0.16\fermi^{-3}$.

Although none of the published microscopic calculations of the critical
temperature $T_c$ has presented a convenient fitting formula in terms of
density, the critical temperatures are roughly quadratic functions of
the Fermi wavevector $k_{\{np\}}= (3\pi^2 n_{\{np\}})^{1/3}$.  I
therefore use
\begin{equation}
\label{eq:Tc}
   T_c(k)=T_{c0} \left[1-\frac{(k-k_0)^2}{(\Delta_k/2)^2} \right]
\end{equation}
as the functional form of $T_c$ for the proton $^1S_0$, neutron $^1S_0$,
and neutron $^3P_2$ states.  The parameters $T_{c0}$, $k_0$, and
$\Delta_k$ are chosen (see Table \ref{t:superfluid}) to approximate the
transition temperature of \citet{amundsen85:triplet} for the neutron
$^3P_2$ state and the transition temperature of
\citet{amundsen85:singlet} for the proton and neutron singlet states.
This choice of $T_c$ reflects the calculations of \citet{takatsuka93} as
well.

\begin{deluxetable}{llll}
\tablewidth{0pt}
\tablecaption{\label{t:superfluid}%
   Parameters used to calculate superfluid transition temperature
   (eq.~[\protect\ref{eq:Tc}]).
}
\tablehead{%
   \colhead{State} & 
   \colhead{$T_c$} & 
   \colhead{$k_0$} & 
   \colhead{$\Delta k$} \\
   &
   \colhead{($\MeV$)} &
   \colhead{($\fermi^{-1}$)} &
   \colhead{($\fermi^{-1}$)}
}
\startdata
proton $^1S_0$ & 0.345 & 0.7 & 1.0\\
neutron $^1S_0$ & 0.802 & 0.7 & 1.2\\
neutron $^3P_2$ & 0.0776 & 2.0 & 1.6
\enddata
\end{deluxetable}

\begin{figure}[tp]
\centering{\includegraphics[width=88mm]{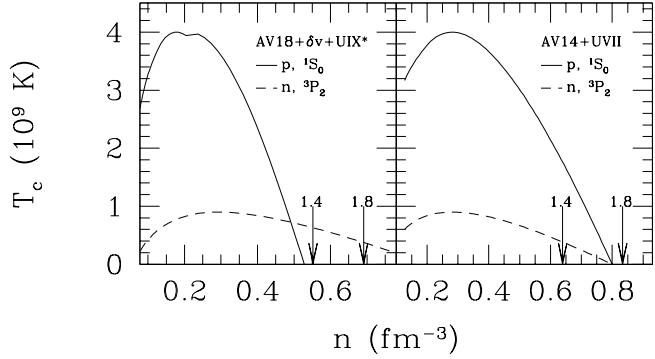}}
\caption{\label{fig:Tc}%
   Superfluid transition temperature $T_c$ (see
   eq.~[\protect\ref{eq:Tc}]) as a function of density, for both the
   proton $^1S_0$ (\emph{solid lines}) and neutron $^3P_2$ (\emph{dashed
   lines}) pairing.  Each panel is for a different core EOS, as
   indicated in the legend.  The arrows mark the central densities, for
   each EOS, of a neutron star of gravitational mass $\Mg=1.4\Msun$ and
   $\Mg=1.8\Msun$.
}
\end{figure}

Figure~\ref{fig:Tc} displays $T_c$ for the proton $^1S_0$ (\emph{solid
lines}) and neutron $^3P_2$ (\emph{dashed lines}) pairing as a function
of density, for both equations of state.  While the neutron critical
temperatures are roughly identical, the proton critical temperature for
\avmod\ vanishes at lower $n$.  This cutoff is because \avmod\ has a
higher proton fraction, at a given $n$, than \avold.  For each EOS, I
show (\emph{arrows}) the central densities of neutron stars of
gravitational masses $\Mg=1.4\Msun$ and $\Mg=1.8\Msun$.  At core
temperatures $\sim 5\ee{8}\K$, each of the four EOS/mass combinations
has normal protons, neutrons, or both in some part of the core.  As a
result, the modified Urca processes still play an important role in the
neutron star's thermal balance\footnote{%
   Recently, neutrino emission from the formation and destruction of
   Cooper pairs has received renewed interest \protect\citep*[see][and
   references therein] {yakovlev99:_neutr_cooper}.  For temperatures
   near the superfluid transition temperature, the neutrino emissivity
   is \emph{enhanced} over that of modified Urca processes.  I have not
   included this emissivity here; for an impure crust, this omission is
   not critical (see \S~\protect\ref{sec:Simple-expr-crust}).}.

\subsection{Heat transport}\label{sec:heat-transport}

Throughout the crust, relativistic electrons transport the heat.  In the
relaxation-time approximation, the conductivity is
\citep[e.g.,][]{ziman72:_princ},
\begin{equation}
\label{eq:conductivity}
   K = \frac{\pi^2}{3} \kB \frac{\kB T n_e}{ m_e^*} \tau.
\end{equation}
Here $m_e^* = \EF/c^2$ is the effective electron mass, and $\tau$ is the
relaxation time.  Where the ions are crystallized, $1/\tau = 1/\tau_{ee}
+ 1/\tau_{eQ} + 1/\tau_{ep}$, and where they are liquefied, $1/\tau =
1/\tau_{ee}+1/\tau_{ei}$.  In these formulae, $\tau_{ee}$, $\tau_{eQ}$,
$\tau_{ep}$, and $\tau_{ei}$ are respectively the relaxation times for
electron-electron (\citealt{urpin80:_therm}; \citealt*{potekhin97}),
electron-impurity \citep{itoh93}, electron-phonon \citep{baiko95}, and
electron-ion \citep{yakovlev80:_therm} scattering.  Electron-electron
scattering is typically negligible over much of the crust because the
strong degeneracy of the electrons restricts the available phase space.

The electron-impurity relaxation time is
\citep{yakovlev80:_therm}
\begin{equation}
\label{eq:eQ-scattering}
   \tau_{eQ} = \frac{p^2_{\rm F} v_{\rm F}}{4\pi Q e^4
   n_N}\Lambda_{eQ}^{-1}.
\end{equation}
Here $p_{\rm F}$ and $v_{\rm F}$ are the momentum and velocity of an
electron at the Fermi surface and $\Lambda_{eQ}\approx 2$
\citep{yakovlev80:_therm} is the logarithmic Coulomb factor.  The
concentration of impurities enters through $Q$, which for a large number
of ion species is \citep{itoh93}
\begin{equation}
\label{eq:impurity}
   Q \equiv \frac{1}{n_N} \sum_i n_i
   \left( Z_i - \langle Z \rangle \right)^2.
\end{equation}
Here $n_i$ and $Z_i$ are the density and charge number of the $i$\/th
species, $n_N=\sum_i n_i$ is the total ionic density, and $\langle Z
\rangle= n_N^{-1} \sum_i n_i Z_i$ is the mean ionic charge number.

There are several contributions to the impurities in the crust.  First,
within each electron capture layer there are at least two species
present.  The electron capture from an even-even to an odd-odd nucleus
is immediately followed by a second electron capture to a lower energy
even-even nucleus \citep{haensel90a}.  Within each capture layer the
impurity parameter is then $Q=4 n_Z n_{Z-2}/(n_Z + n_{Z-2})^2\le 1$,
with $n_Z$ and $n_{Z-2}$ denoting the densities of the two species.
Although \citet{haensel90b,haensel90a} treated the electron captures as
sharp transitions in the crust, in actuality the layers have a finite
thickness set by competition between the flow timescale and the (weak)
electron capture timescale
\citep{bisnovatyi-kogan79:_noneq_x,blaes90:_slowl,bildsten98}.  The
zero-temperature electron capture rate is proportional to
$(\EF-\delta)^3$, where $\delta$ is the reaction threshold.  Where
degenerate electrons supply the pressure, $\EF$ must increase with
depth.  The layers are thin in $\EF$, and because $p\propto\EF^4$, the
layers are geometrically thin as well.  For densities greater than
neutron drip, however, the electrons no longer support the crust, and
$\EF$ need not increase with depth.  In fact, $\EF$ is actually less
following an electron capture layer if the interface between layers of
different composition is treated as a infinitely thin plane.  The
layers, although thin with respect to $\EF$, are then geometrically
thick.  In actuality, thermal broadening of the electron Fermi surface
causes many of the captures to occur pre-threshold, and the capture
layers are thickened to nearly the width between layers
\citep{ushomirsky00:_crust}.  Should the capture layers overlap, then
$Q$ in the mixed layer can become larger than unity.

The impurities manufactured within the capture layers are probably a
small perturbation compared to those already present in the mixture
entering the top of the crust.  \citet{schatz99} found that $Q\sim 100$
immediately following the end of stable hydrogen burning.  An accurate
assessment of $Q$, throughout the crust, requires evolving the
composition of an accreted fluid element on its journey through the
crust.  This task is beyond the scope of this initial survey, and I
instead set upper and lower bounds on the conductivity. The upper bound
to the conductivity is that of a pure crystal (electron-phonon
scattering).  To set the lower bound, first note that electron-impurity
scattering dominates the conductivity wherever $\tau_{eQ}<\tau_{ep}$,
with $\tau_{ep}$ being the electron-phonon relaxation time
\citep{baiko95}
\begin{equation}
\label{eq:ep-scattering}
   \tau_{ep} = \frac{\hbar}{\alpha \kB T}\Lambda_{ep}^{-1}.
\end{equation}
Here $\alpha$ is the fine structure constant, and $\Lambda_{ep}\approx
13$ comes from integrating over the phonon spectrum.
Equations~(\ref{eq:eQ-scattering}) and~(\ref{eq:ep-scattering}) imply
that for $Q \gtrsim 0.66(30\MeV/\EF) (Z/26) (\kB T/0.05\MeV)$
electron-impurity scattering determines the thermal conductivity in the
crust.  If the reactions in the crust do not significantly reduce $Q$
from its large value at the base of the hydrogen/helium burning shell,
then \emph{the heat transport in the crust is controlled by
electron-impurity scattering rather than electron-phonon scattering.}

If $Q$ is very large ($\sim Z^2$), then the impurity relaxation time is
roughly that of electron-ion scattering for a pure crystal
\citep{yakovlev80:_therm},
\begin{equation}
\label{eq:ei-scattering}
   \tau_{ei} = \frac{p^2_{\rm F} v_{\rm F}}{4\pi Z^2 e^4
   n_N}\Lambda_{ei}^{-1},
\end{equation}
with $\Lambda_{ei}=\ln[(2\pi Z/3)^{1/3}\sqrt{1.5+3/\Gamma}]-1$.
Basically, the phonon spectrum is extremely disordered in this case.  I
therefore set a lower limit to the conductivity by using electron-ion
scattering, i.e., by treating the ions as if they were liquefied.  For
consistency, I also use the liquid-state neutrino bremsstrahlung
emissivity \citep{haensel96b} in conjunction with the electron-ion
conductivity.  The impurities in the crust reduce the conductivity but
increase the neutrino emissivity.

In the core, heat is mostly carried by electrons, with neutrons
contributing if they are normal \citep{flowers79}.  I neglect here the
neutron conductivity.  This is a good approximation, as the core is
practically isothermal (see \S~\ref{s:results}).  In evaluating the
electron-proton scattering terms, I used an effective proton mass
$m_p^*=0.7m_p$.  The proton superfluidity both reduces the screening
(increases the scattering) and reduces the proton scattering phase space
(suppresses the scattering); I take these factors into account using the
fits of \citet{gnedin95}.

\subsection{Boundary conditions and method of solution}
\label{sec:Bound-cond-meth}

The first boundary condition is simply $L|_{r=0}=0$.  For a fully
self-consistent solution, the correct second boundary condition is a
relation $L(T)|_{r=R}$, usually obtained from a separate photospheric
calculation.  This is unnecessary, however, when the hydrogen and helium
burn steadily.  The large energy release from this burning determines
the temperature in the outer atmosphere, so that the temperature at the
base of the hydrogen/helium burning shell is a function only of
$\dot{M}$, $\Mg$, and $R$ and may be calculated independently.  In
\S~\ref{s:results}, I show that the luminosity flowing outwards from the
crust is in fact much smaller than that generated by the hydrogen/helium
burning.

The second boundary condition, then, is $T|_{r=R} = T_\circ(\dot{m})$,
where I take $T_\circ$ from \citet{schatz99}.  Here $\dot{m}$ is the
accretion rate per unit area; the fiducial rate used by \citet{schatz99}
is the Eddington rate appropriate for a Newtonian star of $\Mg=1.4\Msun$
and $R=10\km$ accreting a solar composition plasma.  Numerically, this
rate is $\dot{m_E}=8.8\ee{4}\gram\cm^{-2}\second^{-1}$ and is an
excellent approximation to the lowest local accretion rate at which the
hydrogen/helium burning is stable \citep[see,
e.g.,][]{bildsten98:_nuclear}.  At $\dot{m}=\dot{m}_E$, the temperature
at the base of the hydrogen/helium burning shell is $T_\circ=5\ee{8}\K$.
Although the Newtonian surface gravity used by \citet{schatz99} is less
than the values used here, $T_\circ$ is relatively insensitive to $g$
($T_\circ\propto g^{1/7}$; \citealt{bildsten98:_nuclear}), so I do not
adjust it for each model.  This is not critical, as the thermal profile
in the crust is insensitive to $T_\circ$ (see \S~\ref{s:results}).

The accretion rate enters equations~(\ref{eq:entropy})
and~(\ref{eq:flux}) through $\eN$, which is scaled to the accretion
luminosity $L_A$.  Because $T_\circ$ is only a function of $\dot{m}$, I
use the same $\dot{m}$ for each model; the luminosity from this
accretion is then different for each EOS and mass.  The global accretion
rate, as measured by an observer infinitely far away, is
\citep{ayasli82} $\dot{M}=4\pi R^2 \dot{m}/(1+z)$, and the luminosity is
\begin{equation}\label{eq:fiducial-luminosity}
   L_A = \frac{z}{1+z}\dot{M}c^2 = \frac{z}{(1+z)^2} 4\pi R^2
	\dot{m}c^2.
\end{equation}
For the fiducial local accretion rate $\dot{m}_E$, model \modelone\ has
$L_A=\LAo=1.95\ee{38}\erg\second^{-1}$; \modeltwo\ has
$\LAo=2.28\ee{38}\erg\second^{-1}$; \modelthree,
$2.12\ee{38}\erg\second^{-1}$; and \modelfour,
$2.51\ee{38}\erg\second^{-1}$.  To solve the thermal structure,
equations~(\ref{eq:entropy}) and~(\ref{eq:flux}) are finite-differenced
onto the mesh defined by the integration of
equations~(\ref{e:radius})--(\ref{e:pressure}).  An initial guess is
constructed by fixing the temperature throughout the star to $T_\circ$
and integrating equation~(\ref{eq:entropy}) from $L|_{r=0}=0$.  This
trial guess is then iteratively refined by a relaxation technique
\citep{pre92}.  The resolution of the mesh was tested by computing
models with step fractions (see eq.~[\ref{eq:stepsize}]) $f=0.05$ and
$f=0.02$.

\section{Results}\label{s:results}

\subsection{The influence of the microphysics on the thermal profile}
\label{sec:infl-micr-therm}

\begin{figure}[tp]
\centering{\includegraphics[width=88mm]{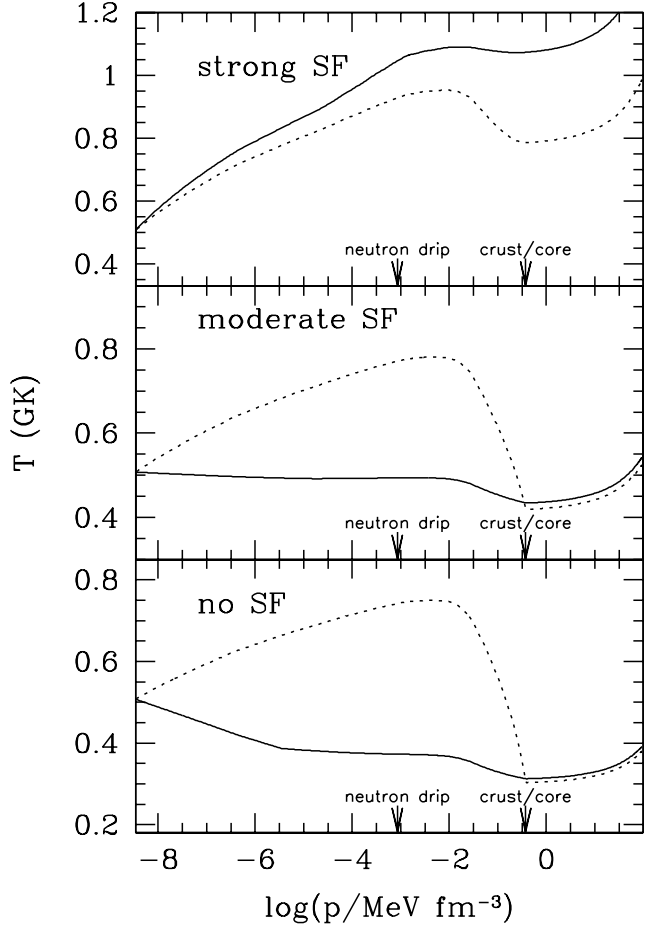}}
\caption{\label{fig:pQ}%
   Proper temperature, in units of $10^9\K$, as a function of pressure.
   The hydrostatic structure is \modelthree\ (see
   Table~\protect\ref{t:models}).  Two cases are compared, with the
   conductivity dominated either by electron-phonon scattering
   (\emph{solid lines}) or by electron-ion scattering (\emph{dotted
   lines}).  The three panels, from top to bottom, show the variation of
   the thermal profile with core superfluidity.  In the top panel, the
   core neutrino emission is completely suppressed: both protons and
   neutrons are superfluid throughout the core.  The middle panel shows
   the effects of moderate superfluidity (see
   Table~\protect\ref{t:superfluid}), whereas in the bottom panel there
   is no superfluidity.  Note that for a low thermal conductivity, the
   maximum crust temperature ($\approx 8\ee{8}\K$) varies only slightly
   with the core temperature.  The pressures of neutron drip and the
   crust-core boundary (see Table \protect~\ref{t:models}) are marked
   along the bottom axis of each plot with arrows.
}
\end{figure}

As promised in section~\ref{sec:heat-transport}, I survey the
uncertainties in the thermal conductivity by solving the thermal
structure (eq.~[\ref{eq:entropy}] and eq.~[\ref{eq:flux}]) with the
conductivity alternately set by electron-phonon scattering and
electron-ion scattering.  Figure~\ref{fig:pQ} shows the thermal profiles
for these two cases with different degrees of superfluidity: strong
(both neutrons and protons are superfluid with $T_c\gg T$ throughout the
core; \emph{top panel}), moderate (corresponding to the parameters in
Table~\ref{t:superfluid}; \emph{middle panel}), and nonexistent
($T_c=0\K$ for both neutrons and protons; \emph{bottom panel}).  The
hydrostatic structure used in this plot is model \modelthree\ (see
Table~\ref{t:models}).  I plot the proper temperature (i.e., the
temperature a local thermometer would measure) because it controls the
conductivity and neutrino emissivity.

\begin{figure}[tp]
\centering{\includegraphics[width=88mm]{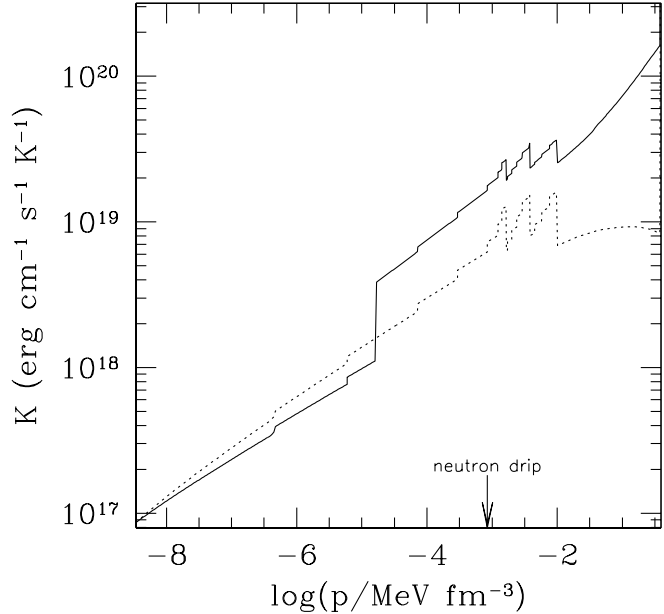}}
\caption{\label{fig:conductivity}%
   Thermal conductivity in the accreted crust.  The thermal structure
   corresponds to the cases shown in the middle panel of
   Figure~\protect\ref{fig:pQ}.  Electron-phonon is denoted with a solid
   line, while electron-ion is denoted by a dotted line.  Note that
   there is an order of magnitude difference between the two cases in
   the inner crust.
}
\end{figure}

The lower conductivity and enhanced bremsstrahlung from electron-ion
scattering (\emph{dotted lines}) produce a greater temperature variation
throughout the crust than if the conductivity and bremsstrahlung were
determined by electron-phonon scattering (\emph{solid lines}).  In the
inner crust, the electron-phonon conductivity
(Fig.~\ref{fig:conductivity}, \emph{solid line}) is an order of
magnitude greater than the electron-ion conductivity
(Fig.~\ref{fig:conductivity}, \emph{dotted line}).  As a result, the
thermal gradient in a locally pure crust is very small, so that the
inner crust temperature is not appreciably different from that of the
core.  A striking result for this case is that the peak crust
temperature depends only weakly on the core temperature.  This is a
consequence of the requirement that a large thermal gradient is needed
to carry the flux in the inner crust when the conductivity is reduced.

\begin{figure}[tp]
\centering{\includegraphics[width=88mm]{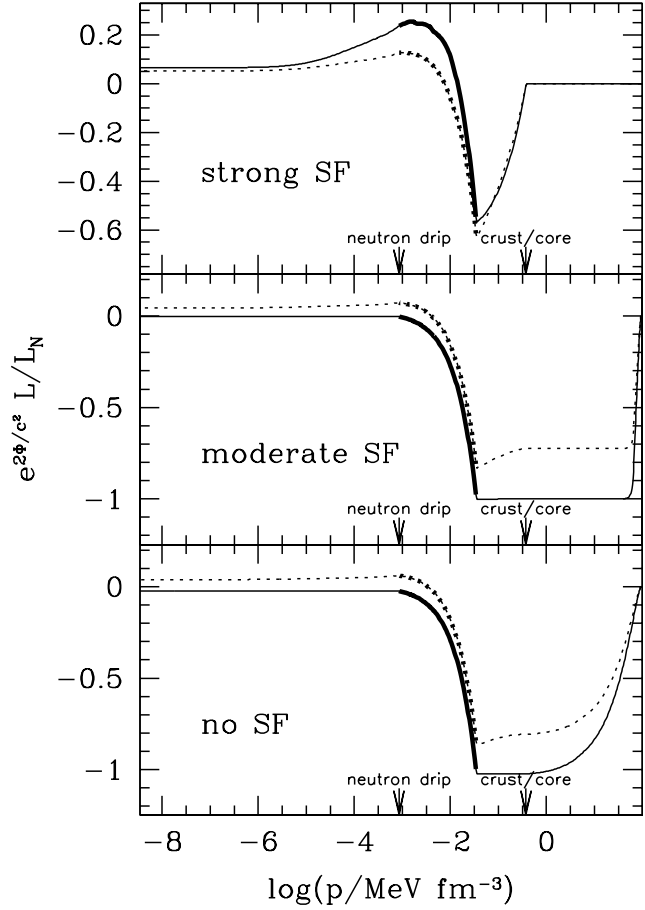}}
\caption{\label{fig:pQ-L}%
   Luminosity, as measured by an infinitely distant observer and in
   units of the total crust nuclear luminosity $L_N$, plotted as a
   function of pressure.  Negative values indicate an inwardly directed
   flux.  The panels correspond to those in Figure
   \protect~\ref{fig:pQ}, and both high conductivity (electron-phonon
   scattering; \emph{solid lines}) and low conductivity (electron-ion
   scattering; \emph{dotted lines}) cases are shown.  The region where
   nuclear heating occurs is indicated with boldfaced lines.
}
\end{figure}

The relative amounts of neutrino emission from the crust and core are
displayed in Figure~\ref{fig:pQ-L}, which shows the luminosity measured
by an observer at infinite distance and scaled to $L_N$.  The lines and
panels correspond to those of Figure~\ref{fig:pQ}, and for reference the
region where nuclear heating occurs (see \S~\ref{sec:nucl-heat-neutr})
is denoted with boldfaced lines.  A negative luminosity indicates that
the heat flow is inward.  When $T_c\gg T$ throughout the core for both
neutrons and protons (\emph{top panel}), all of the neutrinos are
emitted from the crust, and the luminosity is zero throughout the core.
For moderate superfluidity (Table~\ref{t:superfluid}; \emph{middle
panel}), some neutrino emission occurs in the crust when the
conductivity and crust bremsstrahlung are dominated by electron-ion
scattering (\emph{dotted line}), but the bulk of the neutrinos are
emitted in the innermost core, where the protons are normal.  If there
is no superfluidity, then neutrino emission occurs throughout the core
(\emph{bottom panel}).  The decrease in $L$ at pressures $\lesssim
10^{-3}\MeV\fermi^{-3}$ marks where the plasma neutrino emissivity
dominates.

\begin{figure}[tp]
\centering{\includegraphics[width=88mm]{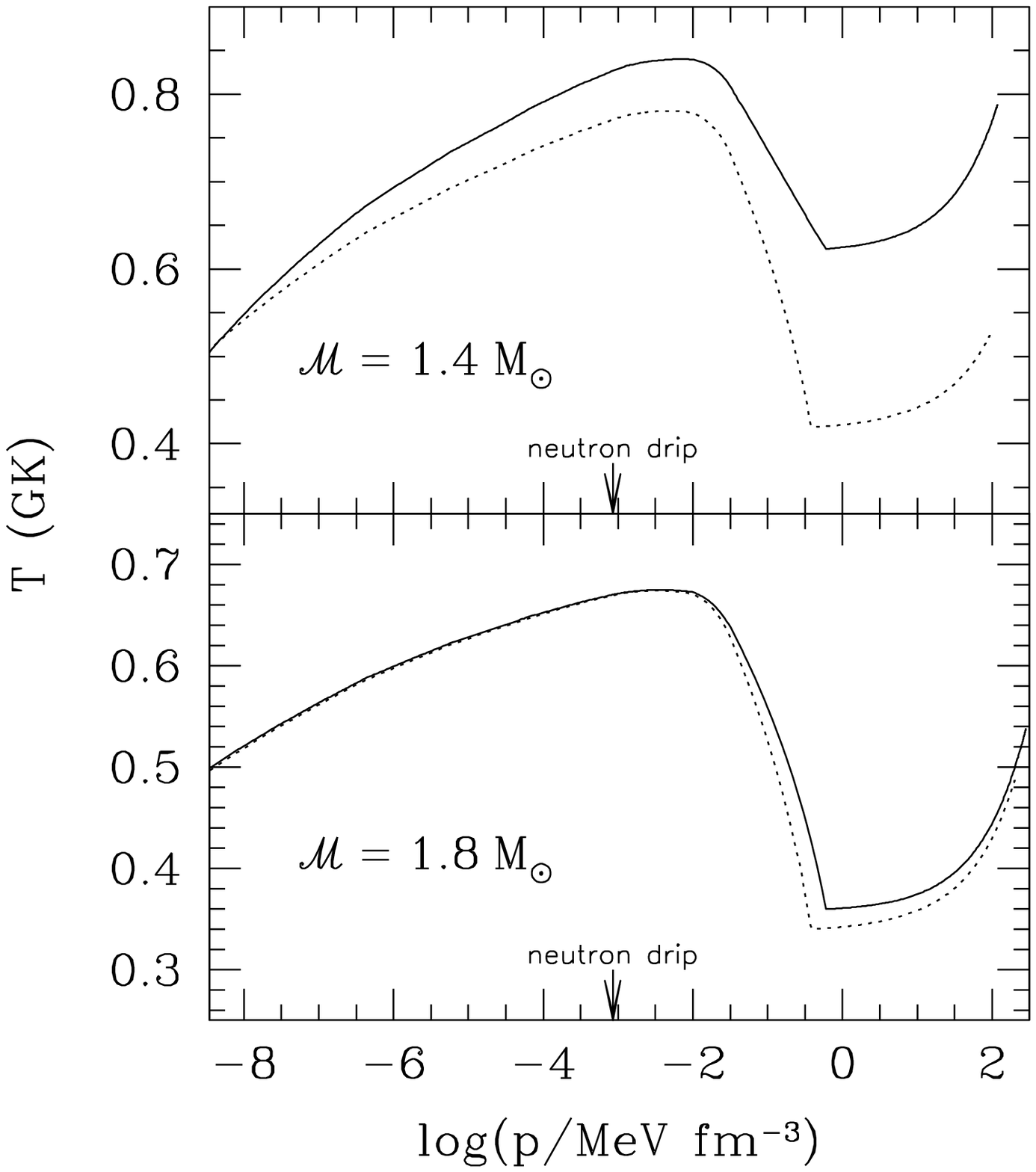}}
\caption{\label{fig:compareEOS}%
   Proper temperature, in units of $10^9\K$, as a function of pressure
   for two different core equations of state, \avold\ (\emph{solid
   lines}) and \avmod\ (\emph{dotted lines}), and two different
   gravitational masses, $\Mg=1.4\Msun$ (\emph{top panel}) and
   $\Mg=1.8\Msun$ (\emph{bottom panel}).  The superfluidity is as
   described in Table~\protect\ref{t:superfluid}, and the conductivity
   is set by election-ion scattering throughout the crust.  The
   pressures of the crust-core boundaries for the two EOS is given in
   Table \protect\ref{t:models}.
}
\end{figure}

\begin{figure}[tp]
\centering{\includegraphics[width=88mm]{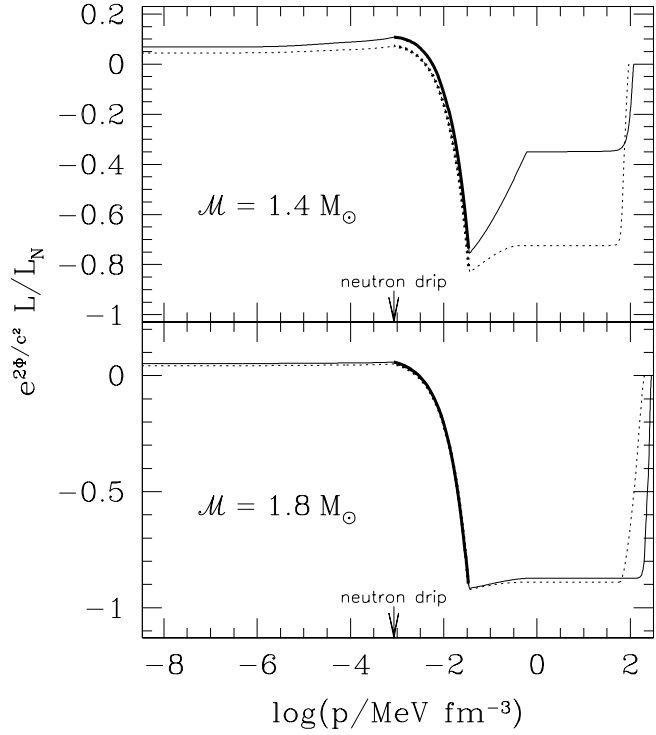}}
\caption{\label{fig:compareEOS_L}%
   Luminosities corresponding to the thermal profiles in
   Fig.~\ref{fig:compareEOS}, with negative values indicating that the
   flux is inwardly directed.  The higher neutrino emissivity from the
   modified Urca proton branch is evident for model \modelthree\
   (\emph{top panel, dotted line}): the \avmod\ EOS has a larger proton
   fraction $Y_p$ and the protons in the inner portion of the core are
   normal (see Fig.~\protect\ref{fig:Tc}).  Nuclear heating occurs in
   the region denoted with boldfaced lines.
}
\end{figure}

The core temperatures of the middle and bottom panels of
Figure~\ref{fig:pQ} are similar because the modified Urca emissivity is
strongly temperature sensitive, so that the core temperature need only
slightly adjust to compensate for a reduced normal core fraction.  To
demonstrate this further, Figures~\ref{fig:compareEOS}
and~\ref{fig:compareEOS_L} show, as functions of pressure, the proper
temperatures and luminosities for the four models in
Table~\ref{t:models} with conductivity set by electron-ion scattering.
The equations of state \avold\ (\emph{solid lines}) and \avmod\
(\emph{dotted lines}) are compared for $\Mg=1.4\Msun$ (\emph{top panel})
and $\Mg=1.8\Msun$ (\emph{bottom panel}).  For $\Mg=1.8\Msun$ the two
equations of state have similar thermal profiles because both have
normal protons and neutrons in at least some fraction of the core (see
Figure~\ref{fig:Tc}).  In contrast, for $\Mg=1.4\Msun$, only
\modelthree\ (\emph{top panel, dotted line}) has normal protons in its
innermost core, and so its core temperature is quite cooler than that of
\modelone\ (\emph{top panel, solid line}).  As a result, \modelone\ has
a stronger neutrino emission from the crust
(Figure~\ref{fig:compareEOS_L}, \emph{top panel, solid line}).  As in
Figure~\ref{fig:pQ-L}, the nuclear heating region is denoted with
boldfaced lines.

\begin{figure}[tp]
\centering{\includegraphics[width=88mm]{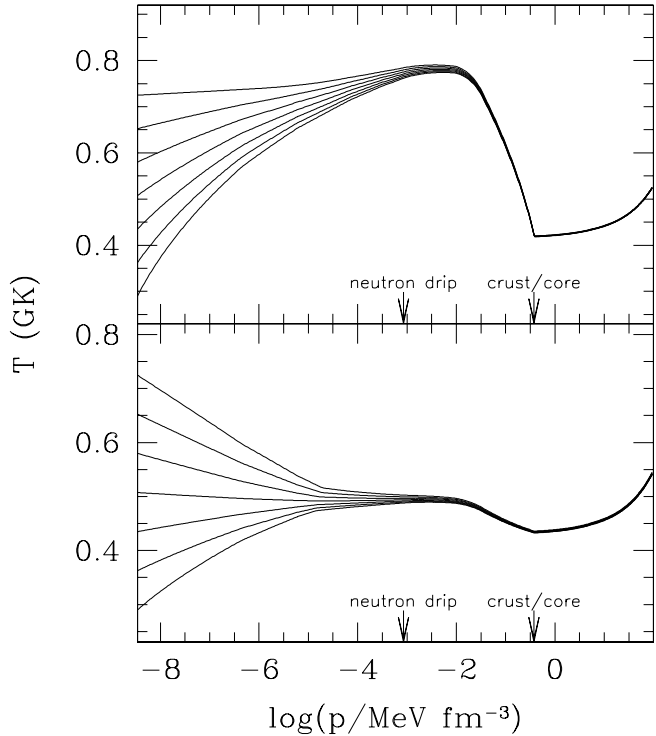}}
\caption{\label{fig:outer-boundary}%
   Proper temperature, in units of $10^9\K$, as a function of pressure.
   A family of solutions for different $T_\circ$, the temperature at the
   top of the crust, is shown, both for a thermal conductivity set by
   electron-ion scattering (\emph{top panel}) and electron-phonon
   scattering (\emph{bottom panel}).  The temperature sensitivity of
   modified Urca processes (superfluidity as described in
   Table~\protect\ref{t:superfluid}) keeps the base of the crust at a
   temperature $\approx 4.3\ee{8}\K$ in both cases.
}
\end{figure}

A generic feature of these solutions is that almost all of the nuclear
heat released in the inner crust flows inward and is balanced by
neutrino cooling from either the crust or core.  Only a small amount
($\lesssim 5\%$) of $L_N$ is conducted to the surface.  As a result, the
temperature in the inner crust and core is set by the processes in the
inner crust.  In particular, \emph{the temperature of the inner crust is
nearly independent of the temperature in the hydrogen/helium burning
shell}.  This is explicitly shown in Figure~\ref{fig:outer-boundary},
where I plot the thermal structure for model \modelthree\ but with the
outer boundary temperature allowed to vary.  The luminosity is fixed at
$\LAo=2.12\ee{38}\erg\second^{-1}$.  The top panel shows the case of
electron-ion conductivity; the bottom, electron-phonon.

\subsection{Analytical expressions for the crust temperature}
\label{sec:Simple-expr-crust}

\newcommand{\ND}{\mathrm{ND}}

The numerical results presented above can be easily understood by
considering a crude analytical calculation of the thermal structure.
The approach is similar to that of \citet{hernquist84:_analy}, with
three exceptions: I include heating from the crust reactions, I fit the
pressure as a function of density, rather than presume a degenerate
relativistic EOS, and I assume the conductivity is determined by
electron-ion, rather than electron-phonon scattering.

In the crust, the surface gravity and radius are roughly constant, and
the differential expressions for the radius (eq.~[\ref{e:radius}]),
pressure (eq.~[\ref{e:pressure}]) and flux (eq.~[\ref{eq:flux}]) can be
combined into the plane-parallel Newtonian equation
\begin{equation}
\label{eq:flux-Newtonian}
    g\rho K \frac{dT}{dp} = \frac{L}{4\pi R^2}.
\end{equation}
To construct this analytical model, I consider the heating and cooling
emissivities to be $\delta$-functions.  Since the flux is then constant
between points where these sources or sinks reside, I may integrate
equation~(\ref{eq:flux-Newtonian}) piecewise between these points, with
$L$ stepping discontinuously at each point \citep{brown98a}.

Integrating equation~(\ref{eq:flux-Newtonian}) requires a relation
$p(\rho)$.  I approximate the mass density $\rho$ by $m_u\,n$ and fit
the pressure with power-laws in both the electron-dominated and
neutron-dominated regions,
\begin{equation}
\label{eq:eos-fit}
   p = \cases{%
	2.67\ee{30}\rho_{12}^{1.27}\dyne\cm^{-2}, &
		$\rho_{12}\le 0.66$, \cr
	4.97\ee{29}\rho_{12}^{1.42}\dyne\cm^{-2}, &
 		$\rho_{12}\ge 8.9$.\cr }
\end{equation}
Here $\rho_{12}=\rho/10^{12}\gram\cm^{-3}$, and the error in $p$ is less
than 9\% and 5\% for the two density regimes respectively\footnote{%
   In this section I use cgs units for easy comparison with the
   astrophysical literature.}.
The exponent in the electron-dominated regime is less than $4/3$ because
the fit accounts for the decrease in $Y_e$ by electron captures.  For
densities just above neutron drip ($0.6607<\rho_{12}<8.913$), the
pressure cannot be fit by a simple power-law in density (see
Figure~\ref{fig:crustEOS}).  This region is where most of the heat is
released, and so for simplicity I presume the region to be isothermal
and place the heating $\delta$-function inside it.

Inserting the expression for the electron-ion scattering frequency
(eq.~[\ref{eq:eQ-scattering}]) into the expression for the thermal
conductivity (eq.~[\ref{eq:conductivity}]) and expanding, I have
\begin{eqnarray}
   K &\approx& \frac{1}{8} \left(\frac{\pi^5}{9}\right)^{1/3}
      \alpha^{-2}\hbar^{-1} \frac{Y_e^{1/3}}{Z}
      \left(\frac{\rho}{m_n}\right)^{1/3} \kB^2 T \nonumber \\
   &\approx&  1.16\ee{20} \left(\frac{Y_e^{1/3}}{Z}\right)
      \rho_{12}^{1/3} T_9 \erg\cm^{-1}\second^{-1}\K^{-1}.
\label{eq:fit-K}
\end{eqnarray}
In this expression, I set the Coulomb logarithm $\Lambda_{ei}$ to unity
and use the shorthand $T_9=T/10^9\K$.  The composition enters through
$Y_e^{1/3}/Z$; for densities less than neutron drip I use
$Y_e^{1/3}/Z=0.0298$, appropriate for a pure iron composition, and for
densities greater than neutron drip I use $Y_e^{1/3}/Z=0.0167$, as
follows from the last entry of Table 2 in \citet{haensel90a}.  (For
impurity scattering, $Y_e^{1/3}/Z \to Z Y_e^{1/3}/Q$.)  Using
equations~(\ref{eq:eos-fit}) and~(\ref{eq:fit-K}), I integrate
equation~(\ref{eq:flux-Newtonian}) from the top of the crust to neutron
drip, $10^{-5}<\rho_{12}<0.66$, and over the inner crust to the core,
$8.913<\rho_{12}<166$, to obtain
\begin{equation}
\label{eq:T-integrand-low}
   T_{9}^2(\rho_{12}) = T_{\ND,9}^2 - 1.30 L_{o,35} R_{10}^{-2}
   g_{14.3}^{-1} \left(\rho_{12}^{-0.060}-1.03\right)
\end{equation}
for $\rho_{12} \le 0.66$ and
\begin{equation}
\label{eq:T-integrand-high}
   T_{9}^2(\rho_{12}) = T_{\ND,9}^2 + 0.33 L_{i,35} R_{10}^{-2}
   g_{14.3}^{-1} \left(\rho_{12}^{0.087}-1.21\right)
\end{equation}
for $\rho_{12}\ge 8.9$.  In these equations $T_\ND$ is the temperature
at neutron drip (presumed constant for $0.66<\rho_{12}<8.9$),
$R_{10}=R/10\km$, $g_{14.3}=g/10^{14.3}\cm\second^{-2}$, and $L_{o,35}$
and $L_{i,35}$ are the luminosities, in units of
$10^{35}\erg\second^{-1}$, for $\rho_{12}<0.66$ and $\rho_{12}>8.9$,
respectively.  Both $L_o$ and $L_i$ are signed: they are positive if the
flux is directed outwards and negative if directed inwards.  Notice that
the coefficient of $L_{o,35}$ in equation~(\ref{eq:T-integrand-low}) is
an order of magnitude larger than the coefficient of $L_{i,35}$ in
equation~(\ref{eq:T-integrand-high}).  This disparity reflects that the
inner crust requires a much smaller thermal gradient than the outer
crust to carry a given flux.

To solve for the thermal structure, I also require that the luminosity
flowing away from the crust heat source is
\begin{equation}\label{eq:luminosity-sum}
   L_N=L_o-L_i,
\end{equation}
and that the core neutrino luminosity balances the heat conducted into
the core,
\begin{equation}\label{eq:core-luminosity}
   L_i+L_\nu(\Tcore)=0.
\end{equation}
Evaluating equation (\ref{eq:T-integrand-low}) at $\rho_{12}=10^{-5}$
and equation~(\ref{eq:T-integrand-high}) at $\rho_{12}=166$, and using
equations~(\ref{eq:luminosity-sum}) and (\ref{eq:core-luminosity}) to
replace $L_o$ and $L_i$ with $L_N$ and $L_\nu$, I obtain an equation for
the core temperature,
\begin{equation}
\label{eq:core-temperature}
   \GTcore^2 = T_{\circ,9}^2 + R_{10}^{-2} g_{14.3}^{-1} 
      \left[ 1.26 L_{N,35} - 1.38 L_{\nu,35}(\GTcore) \right].
\end{equation}
Here $T_\circ=T|_{\rho_{12}=10^{-5}}$.  Solving
equation~(\ref{eq:core-temperature}) for a modified Urca luminosity
$L_{\nu,35}\approx 5\ee{4} \GTcore^8$ and $L_N\approx
1.07\ee{36}\erg\second^{-1} (\dot{m}/\dot{m}_E)$ gives $\GTcore=0.34$
and $L_{\nu,35}=0.92L_{N,35}$, i.e., 92\% of the heat generated in the
crust flows into the core.  This compares reasonably well with the
numerical calculation without core superfluidity (Figures~\ref{fig:pQ}
and~\ref{fig:pQ-L}, \emph{bottom panels}).  In that case, the proper
temperature at the crust bottom is $3.1\ee{8}\K$, the luminosity flowing
out the top of the crust is $0.04L_N$, and the luminosity flowing into
the core is $0.8L_N$.  Neutrino emission from the crust balances the
remainder of $L_N$.

Substituting $\GTcore$ and $L_{\nu,35}$ into
equation~(\ref{eq:T-integrand-high}), I find that $T_{\ND,9}=1.1$, which
is an overestimation of the maximum crust temperature.  This is a
consequence of the ``two-zone'' treatment, which puts all of the
neutrino cooling in the core.  Still, the qualitative features of the
numerical solutions are reproduced.  In
equation~(\ref{eq:T-integrand-high}), the increase in temperature from
core to neutron drip (second term, right-hand side) is much larger than
$\GTcore^2$.  As a result, changing $\GTcore$ has only a small effect on
$T_{\ND,9}$.  Even if the direct Urca process were to operate and cool
the core to $\GTcore \ll 1$, the temperature around neutron drip will
remain high.  \emph{At high accretion rates, the temperature around
neutron drip, for a very impure crust, is primarily determined by the
ability of the inner crust to carry the nuclear luminosity inward and
not so much by the efficiency of core neutrino cooling.}

\subsection{Accretion at higher and lower rates}
\label{sec:approach-to-lower}

\begin{figure}[tp]
\centering{\includegraphics[width=88mm]{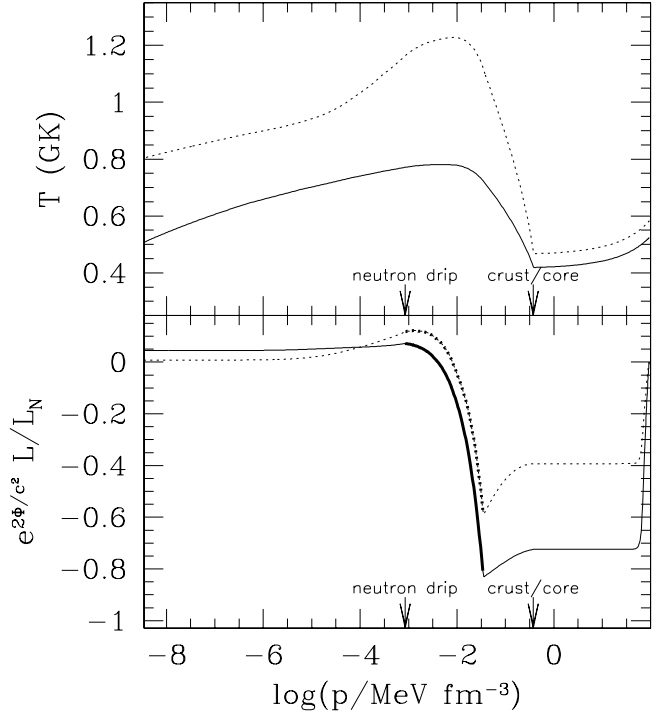}}
\caption{\label{fig:compare-mdot}%
   Proper temperature, in units of $10^9\K$ (\emph{top panel}), and
   luminosity, scaled to the nuclear luminosity, as measured by an
   observer at infinity (\emph{bottom panel}) for accretion luminosities
   $\LAo=2.12\ee{38}\erg\second^{-1}$ (\emph{solid lines}) and $5
   \LAo=1.06\ee{39}\erg\second^{-1}$ (\emph{dotted lines}).  At the
   higher accretion rate, the cooling from crust bremsstrahlung and
   plasma neutrino emission balance the heating (\emph{boldfaced lines})
   on the spot, so that the luminosity flowing out the top of the crust
   is nearly zero.
}
\end{figure}

As the crust temperature increases, crust neutrino bremsstrahlung and
the plasma neutrino process become increasingly important.  At the
higher accretion rate, the brighter crust neutrino luminosity balances
the nuclear heating ``on the spot.''  Figure~\ref{fig:compare-mdot}
compares proper temperature (\emph{top panel}) and scaled luminosity
(\emph{bottom panel}), as measured by an observer at infinite distance,
for model \modelthree\ accreting at $\dot{m}_E$
($L_A=\LAo=2.12\ee{38}\erg\second^{-1}$; \emph{solid line}) and
$5\dot{m}_E$ ($L_A=5\LAo=1.06\ee{39}\erg\second^{-1}$; \emph{dotted
line}).  The conductivity in both cases is set by electron-ion
scattering.  As the crust neutrino cooling increases, a smaller fraction
of $L_N$ flows outward from the top of the crust.

\begin{figure}[tp]
\centering{\includegraphics[width=88mm]{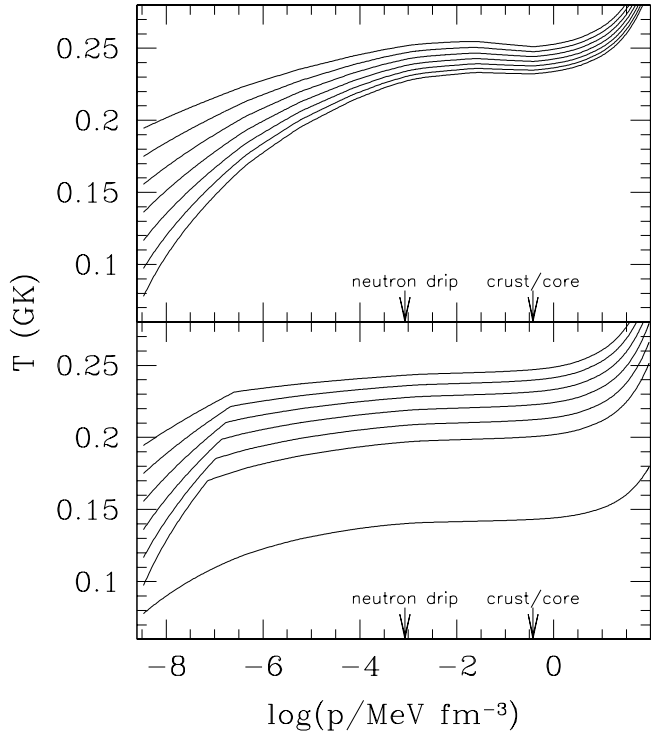}}
\caption{\label{fig:low-mdot-sf}%
   Temperature, in units of $10^9\K$, as a function of pressure for an
   accretion luminosity 1/100 of the fiducial rate plotted in
   Fig.~\protect\ref{fig:outer-boundary}.  Both the high-conductivity
   (electron-phonon scattering; \emph{bottom panels}) and
   low-conductivity (electron-ion scattering; \emph{top panels}) cases
   are considered.  Unlike Fig.~\ref{fig:outer-boundary}, the
   temperature in the inner crust and core depends more on the
   temperature at the top of the crust.  The sharp kink in the thermal
   profiles (\emph{bottom panel}) is where the ions crystallize.
}
\end{figure}

At lower accretion rates, the change in temperature over the inner crust
becomes smaller relative to the core temperature.  The crust becomes
more nearly isothermal and hence more sensitive to the temperatures at
its boundaries \citep[cf.~][]{miralda-escude90,zdunik92}.  From
equation~(\ref{eq:T-integrand-high}), the temperature increase over the
inner crust is $<0.5 T_{\rm core}$ for $L_A<0.06\LAo$, assuming that
$L_i=L_N$.  To demonstrate this, Figure~\ref{fig:low-mdot-sf} displays,
for an accretion luminosity $L_A=0.01\LAo=2.12\ee{36}\erg\second^{-1}$,
the proper temperature and luminosity.  The hydrostatic structure is
\modelthree, the same as in Figure~\ref{fig:outer-boundary}.  The top
panel is for a conductivity set by electron-ion scattering; the bottom
panel is for a conductivity set by electron-phonon scattering.
Solutions for several $T_\circ$ are shown; the range of values are
reduced from those used in Figure~\ref{fig:outer-boundary} by
$(\dot{m}/\dot{m}_E)^{2/7}=0.01^{2/7}$, which is roughly how the
temperature at the base of a hydrogen/helium burning shell scales with
accretion rate \citep{schatz99}.  Of course, the hydrogen and helium
ignition is unstable in an envelope this cold \citep[see][and references
therein]{bildsten98:_nuclear}, and so $T_\circ$ is determined by the
compression of matter in the atmosphere and by the flux flowing out the
top of the crust.  As $T_\circ$ is reduced, more and more of the heat
generated in the crust flows outwards rather than into the core.  As
found by \citet{zdunik92}, an enhanced core neutrino emissivity will
drastically lower the crust temperature for low accretion rates.

\begin{figure}[tp]
\centering{\includegraphics[width=88mm]{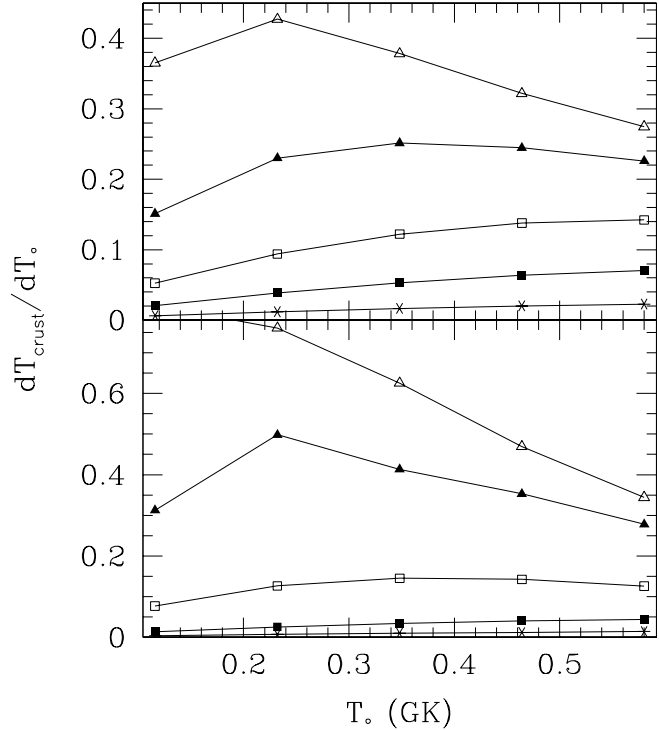}}
\caption[Sensitivity of crust temperature to atmosphere
temperature]{\label{fig:sensitive}%
   Derivative of the crust temperature (at the centroid of the
   heat-producing region, $p=0.017\MeV\fermi^{-3}$) with respect to the
   temperature at the top of the crust (in units of $10^9\K$) for
   $L_A/\LAo = 0.01$ (\emph{hollow triangles}), 0.03 (\emph{filled
   triangles}), 0.1 (\emph{hollow squares}), 0.3 (\emph{filled
   squares}), and 1.0 (\emph{asterisks}).  The top panel is for a low
   conductivity (electron-ion scattering) crust, while the bottom is for
   a high conductivity (electron-phonon scattering) crust.
}
\end{figure}

To illustrate how the crust temperature changes with the temperature in
the hydrogen/helium burning region ($T_\circ$), I compute the derivative
$dT_\mathrm{crust}/dT_\circ$, where $T_\mathrm{crust}$ is the
temperature at the centroid of the heat-producing region,
$p=0.017\MeV\fermi^{-3}$.  Figure~\ref{fig:sensitive} displays
$dT_\mathrm{crust}/dT_\circ$ as a function of $T_\circ$ for five
different accretion rates: $L_A/\LAo = 0.01$ (\emph{hollow triangles}),
0.03 (\emph{filled triangles}), 0.1 (\emph{hollow squares}), 0.3
(\emph{filled squares}), and 1.0 (\emph{asterisks}).  When the
conductivity is low (electron-ion scattering; \emph{top panel}),
$T_\mathrm{crust}$ is generally less sensitive to $T_\circ$ than when
electron-phonon scattering sets the heat transport (\emph{bottom
panel}).  The derivative (at a given accretion rate) is largest when
$T_\circ=T_\mathrm{crust}$; this peak is evident in the top panel for
$L_A/\LAo=0.01$ and in the bottom panel for $L_A/\LAo=0.03$.  The rapid
rise of $dT_\mathrm{crust}/dT_\circ$ in the bottom panel for $L_A/\LAo =
0.01$ is because the neutrino cooling in the crust and core goes to
zero, so that all of the heat generated in the crust flows outwards
(cf.\ Figure~\ref{fig:low-mdot-sf}, \emph{bottom panel}).  In addition,
the crust in the entire region considered is also crystalline, which
reduces $dT/dp$.  In general, for $\dot{M}\gtrsim
10^{-9}\Msun/\yr^{-1}$, the temperature in the crust becomes independent
of the temperature in the atmosphere and upper ocean of the neutron
star.

\subsection{Crust melting}
\label{sec:Crust-melting}

\begin{figure}[tp]
\centering{\includegraphics[width=88mm]{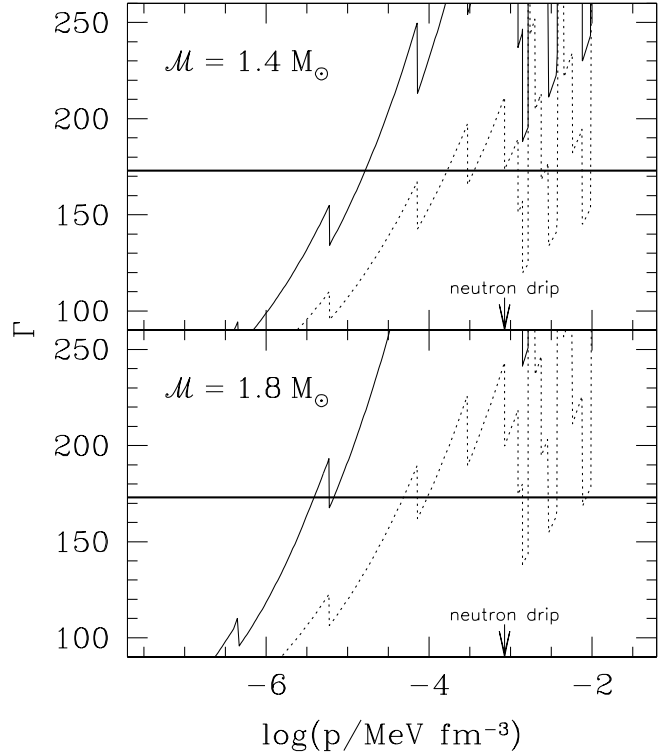}}
\caption{ \label{fig:Gamma}%
   Ion coupling parameter $\Gamma$ as a function of pressure in the
   neutron star crust, for \modelthree\ (\emph{top panel}) and
   \modelfour\ (\emph{bottom panel}), as described in
   Table~\protect\ref{t:models}.  The conductivity is alternately
   dominated by electron-phonon scattering (\emph{solid lines}) and
   electron-ion scattering (\emph{dotted lines}).  I also display the
   melting criterion $\Gamma=173$ (\emph{heavy solid lines}).
}
\end{figure}

An interesting possibility for a rapidly accreting neutron star is that
its crust may melt.  This happens wherever $\Gamma\lesssim 170$, where
the exact value is uncertain \citep[for a review,
see][]{ichimaru82:_stron}.  Since I use the formulation of
\citet{farouki93} to calculate the ionic free energy, I also adopt their
melting value, $\Gamma_M = 173$.  The crust reactions reduce $Z$ and
heat the crust; both of these effects decrease $\Gamma$, as shown in
Figure~\ref{fig:Gamma} for models \modelthree\ (\emph{top panel}) and
\modelfour\ (\emph{bottom panel}), with each model accreting at its
fiducial rate.  In both cases, the core superfluidity is as described in
Table~\ref{t:superfluid}.  For a low thermal conductivity (\emph{dotted
lines}), the crust melts in a series of layers.  The jaggedness of
$\Gamma$ is because of the pycnonuclear reactions.  Each one doubles $Z$
and halves $n_N$, so that $\Gamma$ increases by $2^{5/3}$ and the crust
refreezes.  Electron captures then decrease $Z$ and $\Gamma$ until the
crust melts again.

\begin{figure}[tp]
\centering{\includegraphics[width=88mm]{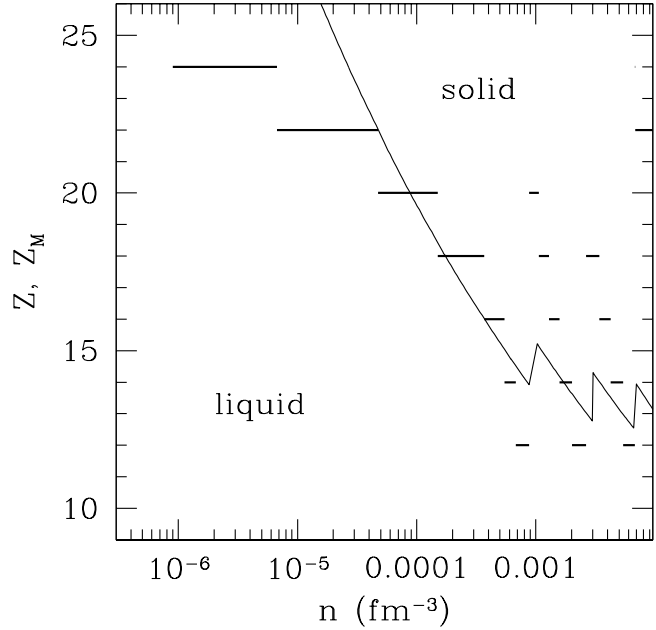}}
\caption{\label{fig:Z-melt}%
   Nuclear charge $Z_M$ (\emph{thin solid line}) such that a pure
   lattice composed of ions of charge $Z<Z_M$ melts, as a function of
   density $n$.  This solution is for the same parameters as
   Fig.~\ref{fig:Gamma}, top panel, dotted line.  The ions are liquid in
   the region below this curve and crystalline in the region above it.
   Also plotted are the $Z$ of the nuclei present at each depth
   (\emph{heavy lines}), according to \protect\citet{haensel90b}.
}
\end{figure}

As a consequence of this melting and freezing, the crust resembles a
layer cake at densities greater than neutron drip.
Figure~\ref{fig:Z-melt} shows the nuclear charge $Z_M$ (\emph{thin
line}), below which the ions are liquid, along with the $Z$ of the
nuclei present (\emph{thick lines}) according to \citet{haensel90b}.
The thermal structure is the same as plotted in Figure~\ref{fig:Gamma},
top panel, dotted line.  $\Gamma$ increases with density (or
equivalently, $Z_M$ decreases), and so the naive expectation is a sharp
transition from an ionic ocean to a crust.  The Fermi energy also
increases with density, however, and the ensuing decrease in $Z$ from
electron captures offsets the rise in $\Gamma$: both $Z_M$ and $Z$
decrease together.  The melting strongly depends on composition: an
increment of $Z$ by 2--3 is enough to keep the crust crystalline
throughout.

Notice from Figure~\ref{fig:Gamma} that there is no crustal melting if
the conductivity is solely determined by electron-phonon scattering, for
models \modelthree\ and \modelfour\ (\emph{solid lines}).  For model
\modelone, the crust melts even if the conductivity is set by
electron-phonon scattering.  For larger masses (models \modeltwo\ and
\modelfour), a high impurity concentration is needed to ensure melting.
This is a consequence of the stronger core neutrino cooling holding the
crust at a slightly lower temperature (cf.\ Figures~\ref{fig:compareEOS}
and \ref{fig:compareEOS_L}).  Of course, when the impurity concentration
is high, the single-species calculation of $\Gamma$
(eq.~[\ref{eq:Gamma}]) is no longer applicable.  Calculations for
binary-ionic mixtures \citep[e.g.,][]{segretain93:_cryst} show that the
melting temperature is lowered below that of the pure phases.  While the
phase diagram of a plasma composed of a large number of species has not
been calculated, it is likely that the $\Gamma_M$ of an impure crust is
lower than that assumed here.  This strengthens the contention that the
impure crust of a rapidly accreting neutron star contains melted layers,
provided that the $Z$ used here \citep[from][]{haensel90b} is roughly
the average charge of the nuclei actually present.  A self-consistent
calculation of the crust composition, and the resulting phase diagram,
is required to conclusively determine if layer cake melting actually
occurs.

\section{Summary and concluding remarks}\label{s:conclusion}

There are three main conclusions presented in this work.  First, for
neutron stars accreting rapidly enough for the accreted hydrogen and
helium to burn stably, most of the heat released in the crust flows into
the core.  As a result, the thermal profile in the inner crust is nearly
independent of the temperature at the top of the crust.  Second, if the
crust lattice is very impure, there is a maximum in temperature at
densities greater than neutron drip, where the heating occurs.  The peak
temperature in the crust in this case is set by the ability of the crust
to carry the generated nuclear luminosity inward from the reaction shell
and is relatively insensitive to the core temperature.  Third, heating
the inner crust to temperatures $\approx 8\ee{8}\K$ might melt the crust
in thin layers where electron captures have reduced the ionic charge.

There are several consequences of these results.  Because a fluid layer
does not support shear stress, the strain in the crust must vanish in
these melt layers.  This will limit the quadrupole that can be induced
by thermal perturbations to the electron capture rate
\citep{bildsten98:gravity-wave} if these captures occur above the melt
layer.  In addition, the fluid layers can dissipate rotational energy,
either through hydrodynamical or magnetohydrodynamical processes, and
thus contribute to balancing the accretion torque acting on the stellar
surface.  The electrical conductivity of an accreted crust is reduced,
both because of crust heating \citep{urpin95,geppert94} and because of
crust impurities \citep{brown98a}.  If the crust is as impure as
considered here, the timescale for Ohmic decay over a pressure
scaleheight is much less (by a factor of 100) than the flow timescale,
for much of the crust.  As a result, the inward advection of magnetic
flux \citep{konar97} is reduced in importance.  Thermomagnetic effects,
such as current drift \citep{geppert94} and the battery effect
\citep*[e.g.,][]{blandford83}, will be comparatively more important,
however, because of the greater thermal gradient.

In recent years, attention has been given to other, more efficient,
cooling mechanisms.  The direct Urca process can operate if the proton
fraction is greater than 0.148 \citep{lattimer91} or if hyperons are
present \citep{prakash92:_rapid_delta}.  Other exotic mechanisms may be
possible, including pion condensates \citep{umeda94}, kaon condensates
\citep{brown88:_stran}, or quark matter \citep{iwamoto82:_neutr}.  The
exotic mechanisms have the same temperature dependence as the direct
Urca ($\propto T^6$) but are weaker.  Although none of the hydrostatic
structures considered in this paper has an interior proton fraction
large enough to activate the direct Urca, some form of enhanced cooling
could operate.  However, the crust temperature would still remain high
(\S~\ref{sec:Simple-expr-crust}) if the crust were very impure.  Direct
observational consequences of the core neutrino emissivity are
unfortunately lacking.  It is only in the cooling after accretion halts
and the crust thermally relaxes (as in the transients;
\citealt*{brown98:transients}) that the mode of core neutrino emissivity
can be investigated.  This is unlike the case of isolated, cooling
neutron stars, for which the core neutrino cooling must be treated
correctly.

The results of this investigation show that the most vexing impediment
to further calculations of the thermal structure of an accreting neutron
star, and hence to a better understanding of the issues raised in this
section, is the need to calculate the composition throughout the crust
for the trajectory in $(n,T)$ space followed by an accreted fluid
element.

\acknowledgements

It is a pleasure to thank Lars Bildsten, Andrew Cumming, Andrew Melatos,
and Greg Ushomirsky for many helpful discussions and for reading drafts
of this work.  I also thank Chris Pethick for suggesting that the nuclei
in the inner crust may remain spherical if the charge is low enough and
the referee for helpful comments on the melting of a multi-species
crystal.  This research was supported by NASA grant NAG5-8658.  EFB is
supported by a NASA GSRP Graduate Fellowship under grant NGT5-50052.

\end{document}